\documentclass[journal]{IEEEtran}

\usepackage{amsmath}
\usepackage{amssymb}
\usepackage{amsfonts}
\usepackage{slashbox}
\usepackage{graphicx}
\usepackage{cite, amsmath, amsfonts, amssymb, psfrag, epsfig,tikz}
\usetikzlibrary{shapes,shadows,arrows}
\newtheorem{definition}{{Definition}}
\newtheorem{theorem}{{Theorem}}
\newtheorem{lemma}{{Lemma}}

\newtheorem{remark}{{Remark}}

\DeclareMathAlphabet{\mathpzc}{OT1}{pzc}{m}{it}
\normalsize

\ifCLASSINFOpdf
\else
\fi

\begin{document}
%
\title{Robust Distributed Compression of Symmetrically Correlated Gaussian Sources}
%
%
%
\author{Yizhong Wang, Li Xie, Xuan Zhang, Jun Chen
}

\maketitle

\begin{abstract}

Consider a lossy compression system with $\ell$ distributed encoders and a centralized decoder. Each encoder compresses its observed source and forwards the compressed data to the decoder for joint reconstruction of the target signals under the mean squared error distortion constraint. It is assumed that the observed sources can be expressed as the sum of the target signals and the corruptive noises,  which are generated independently from two symmetric multivariate Gaussian distributions. Depending on the parameters of such distributions, the rate-distortion limit of this  system is characterized either completely or at least for sufficiently low distortions. The results are further extended to the robust distributed compression setting, where the outputs of a subset of encoders may also be used to produce a non-trivial reconstruction of the corresponding target signals. In particular, we obtain in the high-resolution regime a precise characterization of the minimum achievable reconstruction distortion based on the outputs of $k+1$ or more encoders when every $k$ out of all $\ell$ encoders are operated collectively in the same mode that is greedy in the sense of minimizing the distortion incurred by the reconstruction of the corresponding $k$ target signals with respect to the average rate of these $k$ encoders.
\end{abstract}


%
\IEEEpeerreviewmaketitle

\section{Introduction}
%
%
%
%
\IEEEPARstart{C}{onsider} a wireless sensor network where potentially noise-corrupted signals are collected and forwarded to a fusion center for further processing. Due to the communication constraints, it is often necessary to reduce the amount of the transmitted data by local pre-processing at each sensor. Though the multiterminal source coding theory, which aims to provide a systematic guideline for the implementation of such pre-processing, is far from being complete, significant progress has been made over the past few decades, starting from the seminal work by Slepian and Wolf on the lossless case \cite{SW73} to the more recent results on the quadratic Gaussian case \cite{Oohama97, Oohama98, PTR04, CZBW04, Oohama05, CB08, WTV08, TVW10, WCW10, YX12, YZX13, WC13, WC14, Oohama14, CEK17,  CXCWW17}.
Arguably the greatest insight offered by this theory is that one can capitalize on the statistical dependency among the data at different sites to improve the compression efficiency even when such data need to be compressed in a purely distributed fashion. However, this performance improvement comes at a price: the compressed data from different sites might not be separably decodable, instead they need to be gathered at a central decoder for joint decompression. As a consequence, losing a portion of distributedly compressed data may render the remaining portion completely useless. Indeed, such situations are often encountered in practice. For example, in the aforementioned wireless sensor network, it could happen that the fusion center fails to gather the complete set of compressed data needed for performing joint decompression due to unexpected sensor malfunctions or undesirable channel conditions. A natural question thus arises whether a system can harness the benefits of distributed compression without jeopardizing its functionality in adverse scenarios.  Intuitively, there exists a tension between compression efficiency and system robustness. A good distributed compression system should strike a balance between these two factors.
The theory intended to characterize the fundamental tradeoff between compression efficiency and system robustness for the centralized setting is known as multiple description coding, which has been extensively studied \cite{Ozarow80, EGC82, VKG03, PPR04, PPR05, CTBH06, WV07, TCD08, WV09, CTD09, Chen09, TC10, CDZW10, WCZCP11, CZD12, ZDCS12, FWSC13, SSC14, XCW17}. In contrast, its distributed counterpart is far less developed, and the relevant literature is rather scarce \cite{IPRP05, CB082, CW08}.

In the present work we consider a lossy compression system with $\ell$ distributed encoders and a centralized decoder. 
Each encoder compresses its observed source and forwards the compressed data to the decoder. Given the data from an arbitrary subset of encoders, the decoder is required to reconstruct the corresponding target signals within a prescribed mean squared error distortion threshold (dependent on the cardinality of that subset). It is assumed that the observed sources can be expressed as the sum of the target signals and the corruptive noises,  which are generated independently from two (possibly different) symmetric\footnote{This symmetry assumption is not essential for our analysis. It is adopted mainly for the purpose of making the rate-distortion expressions as explicit as possible.} multivariate Gaussian distributions. This setting is similar to that of the robust Gaussian CEO problem studied in \cite{IPRP05, CB082}. However, there are two major differences: the robust Gaussian CEO problem imposes the restrictions that 1) the target signal is a scalar process, and 2) the noises across different encoders are independent. Though these restrictions could be justified in certain scenarios, they were introduced largely due to the technical reliance on Oohama's bounding technique  for the scalar Gaussian CEO problem \cite{Oohama98, Oohama05}. In this paper we shall tackle the more difficult case where the target signals jointly form a vector process by adapting recently developed analytical methods in Gaussian multiterminal source coding theory \cite{WCW10, WC13,WC14, Oohama14} to the robust compression setting. Moreover, we show that the theoretical difficulty caused by correlated noises can be circumvented through a fictitious signal-noise decomposition of the observed sources such that the resulting noises are independent across encoders. In fact, it will become clear that this decomposition can be useful even for analyzing those distributed compression systems with independent noises. Our main results are summarized below.
\begin{enumerate}
\item For the case where the decoder is only required to reconstruct the target signals based on the outputs of all $\ell$ encoders, the rate-distortion limit is characterized either completely or partially, depending on the parameters of signal and noise distributions,

\item For the case where the outputs of a subset of encoders may also be used to produce a non-trivial reconstruction of the corresponding target signals, the minimum achievable reconstruction distortion based on the outputs of $k+1$ or more encoders is characterized either completely or partially, depending on the parameters of signal and noise distributions,
     when every $k$ out of all $\ell$ encoders are operated collectively in the same mode that is greedy in the sense of minimizing the distortion incurred by the reconstruction of the corresponding $k$ target signals with respect to the average rate of these $k$ encoders.
\end{enumerate}

The rest of this paper is organized as follows.  We state the problem definitions and the main results in Section \ref{sec:definition}. The proof is presented in Section \ref{sec:theorem2}. We conclude the paper in Section \ref{sec:conclusion}.

Notation: The expectation operator, the transpose operator, the
trace operator, and the determinant operator are denoted by $\mathbb{E}[\cdot]$, $(\cdot)^T$ , $\mathrm{tr}(\cdot)$, and $\det(\cdot)$, respectively. A $j$-dimensional all-one row vector is written as $1_j$. We use $\mathrm{diag}^{(j)}(\kappa_1,\cdots,\kappa_{j})$ to represent a $j\times j$ diagonal matrix with diagonal entries $\kappa_1,\cdots,\kappa_j$, and use $Y^n$ as an abbreviation of $(Y(1),\cdots,Y(n))$. For a set $\mathcal{A}$ with elements $a_1<\cdots<a_j$, $(\omega_i)_{i\in\mathcal{A}}$ means $(\omega_{a_1},\cdots,\omega_{a_j})$. The cardinality of a set $\mathcal{S}$ is denoted by $|\mathcal{S}|$. Throughout this paper, the base of the logarithm function is $e$.














\section{Problem Definitions and Main Results}\label{sec:definition}

Let  the target signals $X\triangleq(X_1,\cdots,X_{\ell})^T$ and the corruptive noises $Z\triangleq(Z_1,\cdots,Z_{\ell})^T$ be two mutually independent $\ell$-dimensional ($\ell\geq 2$) zero-mean Gaussian random vectors, and the observed sources $S\triangleq(S_1,\cdots,S_{\ell})^T$ be their sum (i.e., $S=X+Z$).
Their respective covariance matrices are given by
\begin{align*}
&\Gamma_X\triangleq\left(
  \begin{array}{cccc}
    \gamma_X & \rho_X\gamma_X & \cdots &  \rho_X\gamma_X \\
    \rho_X\gamma_X & \ddots & \ddots &   \vdots \\
    \vdots & \ddots & \ddots & \rho_X\gamma_X \\
     \rho_X\gamma_X & \cdots & \rho_X\gamma_X & \gamma_X \\
  \end{array}
\right),\\
&\Gamma_Z\triangleq\left(
  \begin{array}{cccc}
    \gamma_Z & \rho_Z\gamma_Z & \cdots &  \rho_Z\gamma_Z \\
    \rho_Z\gamma_Z & \ddots & \ddots &   \vdots \\
    \vdots & \ddots & \ddots & \rho_Z\gamma_Z \\
     \rho_Z\gamma_Z & \cdots & \rho_Z\gamma_Z & \gamma_Z \\
  \end{array}
\right),\\
&\Gamma_S\triangleq\left(
  \begin{array}{cccc}
    \gamma_S & \rho_S\gamma_S & \cdots &  \rho_S\gamma_S \\
    \rho_S\gamma_S & \ddots & \ddots &   \vdots \\
    \vdots & \ddots & \ddots & \rho_S\gamma_S \\
     \rho_S\gamma_S & \cdots & \rho_S\gamma_S & \gamma_S \\
  \end{array}
\right),
\end{align*}
and satisfy $\Gamma_S=\Gamma_X+\Gamma_Z$.
Moreover, we construct an i.i.d. process $\{(X(t),Z(t),S(t))\}_{t=1}^\infty$ such that the joint distribution of $X(t)\triangleq(X_1(t),\cdots,X_{\ell}(t))^T$, $Z(t)\triangleq(Z_1(t),\cdots,Z_{\ell}(t))^T$, and $S(t)\triangleq(S_1(t),\cdots,S_{\ell}(t))^T$ is the same as that of $X$, $Z$, and $S$ for $t=1,2,\cdots$.

By the eigenvalue decomposition, every $j\times j$ (real) matrix
\begin{align*}
\Gamma^{(j)}\triangleq\left(
  \begin{array}{cccc}
    \alpha & \beta & \cdots &  \beta \\
    \beta & \ddots & \ddots &   \vdots \\
    \vdots & \ddots & \ddots & \beta \\
     \beta & \cdots & \beta & \alpha \\
  \end{array}
\right)
\end{align*}
can be written as
\begin{align}
\Gamma^{(j)}=\Theta^{(j)}\Lambda^{(j)}(\Theta^{(j)})^T,\label{eq:decomposition}
\end{align}
where $\Theta^{(j)}$ is an arbitrary (real) unitary matrix with the first column being $\frac{1}{\sqrt{j}}1^T_j$, and
\begin{align*}
\Lambda^{(j)}\triangleq\mathrm{diag}^{(j)}(\alpha+(j-1)\beta,\alpha-\beta,\cdots,\alpha-\beta).
\end{align*}
For $j\in\{1,\cdots,\ell\}$, let $\Gamma^{(j)}_X$, $\Gamma^{(j)}_Z$, and $\Gamma^{(j)}_S$ denote the leading $j\times j$ principal submatrices of $\Gamma_X$, $\Gamma_Z$, and $\Gamma_S$, respectively; in view of (\ref{eq:decomposition}), we have
\begin{align*}
\Gamma^{(j)}_X=\Theta^{(j)}\Lambda^{(j)}_X(\Theta^{(j)})^T,\\
\Gamma^{(j)}_Z=\Theta^{(j)}\Lambda^{(j)}_Z(\Theta^{(j)})^T,\\
\Gamma^{(j)}_S=\Theta^{(j)}\Lambda^{(j)}_S(\Theta^{(j)})^T,
\end{align*}
where
\begin{align*}
&\Lambda^{(j)}_X\triangleq\mathrm{diag}^{(j)}(\lambda^{(j)}_{X,1},\lambda_{X,2},\cdots,\lambda_{X,2}),\\
&\Lambda^{(j)}_Z\triangleq\mathrm{diag}^{(j)}(\lambda^{(j)}_{Z,1},\lambda_{Z,2},\cdots,\lambda_{Z,2}),\\
&\Lambda^{(j)}_S\triangleq\mathrm{diag}^{(j)}(\lambda^{(j)}_{S,1},\lambda_{S,2},\cdots,\lambda_{S,2})
\end{align*}
with
\begin{align*}
&\lambda^{(j)}_{X,1}\triangleq(1+(j-1)\rho_X)\gamma_X,\\
&\lambda_{X,2}\triangleq(1-\rho_X)\gamma_X,\\
&\lambda^{(j)}_{Z,1}\triangleq(1+(j-1)\rho_Z)\gamma_Z,\\
&\lambda_{Z,2}\triangleq(1-\rho_Z)\gamma_Z,\\
&\lambda^{(j)}_{S,1}\triangleq(1+(j-1)\rho_S)\gamma_S,\\
&\lambda_{S,2}\triangleq(1-\rho_S)\gamma_S.
\end{align*}

Note that $\Gamma_X$, $\Gamma_Z$, and $\Gamma_S$ are positive semidefinite (and consequently are well-defined covariance matrices) if and only if
$\lambda^{(\ell)}_{X,1}\geq 0$, $\lambda_{X,2}\geq 0$, $\lambda^{(\ell)}_{Z,1}\geq 0$, $\lambda_{Z,2}\geq 0$, $\lambda^{(\ell)}_{S,1}\geq 0$, and $\lambda_{S,2}\geq 0$.
Furthermore, we assume that $\gamma_X>0$ since otherwise the target signals are not random. It follows by this assumption that $\gamma_S>0$, $\lambda^{(\ell)}_{X,1}+\lambda_{X,2}>0$, and $\lambda^{(\ell)}_{S,1}+\lambda_{S,2}>0$.

\begin{definition}\label{def:robust}
Given $k\in\{1,\cdots,\ell\}$, a rate-distortion tuple $(r,d_k,\cdots,d_{\ell})$ is said to be achievable if, for any $\epsilon>0$, there exist encoding functions $\phi^{(n)}_i:\mathbb{R}^n\rightarrow\mathcal{C}^{(n)}_i$, $i=1,\cdots,\ell$, such that
\begin{align}
&\frac{1}{k n}\sum\limits_{i\in\mathcal{A}}\log|\mathcal{C}^{(n)}_i|\leq r+\epsilon,\nonumber\\
&\hspace{1in}\mathcal{A}\subseteq\{1,\cdots,\ell\}\mbox{ with }|\mathcal{A}|=k,\label{eq:robustconstraint1}\\
&\frac{1}{|\mathcal{A}|n}\sum\limits_{i\in\mathcal{A}}\sum\limits_{t=1}^n\mathbb{E}[(X_i(t)-\hat{X}_{i,\mathcal{A}}(t))^2]\leq d_{|\mathcal{A}|}+\epsilon,\nonumber\\
&\hspace{1in}\mathcal{A}\subseteq\{1,\cdots,\ell\}\mbox{ with }|\mathcal{A}|\geq k,\label{eq:robustconstraint2}
\end{align}
where $\hat{X}_{i,\mathcal{A}}(t)\triangleq\mathbb{E}[X_i(t)|(\phi^{(n)}_{i'}(S^n_{i'}))_{i'\in\mathcal{A}}]$. 
The set of all such achievable $(r,d_k,\cdots,d_{\ell})$ is denoted by $\mathcal{RD}_k$.
\end{definition}
\begin{remark}\label{remark1}
Due to the symmetry of the underlying distributions, it can be shown via a timesharing argument that $\mathcal{RD}_k$ is not affected if we replace (\ref{eq:robustconstraint1}) with either of the following constraints
\begin{align*}
&\frac{1}{n}\log|\mathcal{C}^{(n)}_i|\leq r+\epsilon,\quad i=1,\cdots,\ell,\\
&\frac{1}{\ell n}\sum\limits_{i=1}^\ell\log|\mathcal{C}^{(n)}_i|\leq r+\epsilon,
\end{align*}
and/or replace (\ref{eq:robustconstraint2}) with either of the following constraints
\begin{align*}
&\frac{1}{n}\sum\limits_{t=1}^n\mathbb{E}[(X_i(t)-\hat{X}_{i,\mathcal{A}}(t))^2]\leq d_{|\mathcal{A}|}+\epsilon,\nonumber\\
&\hspace{1in} \mathcal{A}\subseteq\{1,\cdots,\ell\}\mbox{ with }|\mathcal{A}|\geq k,\\
&\frac{1}{{n\choose j}jn}\sum\limits_{\mathcal{A}\subseteq\{1,\cdots,\ell\}:|\mathcal{A}|=j}\sum\limits_{i\in\mathcal{A}}\sum\limits_{t=1}^n\mathbb{E}[(X_i(t)-\hat{X}_{i,\mathcal{A}}(t))^2]\\
&\leq d_{j}+\epsilon,\quad j=k,\cdots,\ell.
\end{align*}
\end{remark}
\begin{remark}
We show in Appendix \ref{app:dmin} that, for $j=k,\cdots,\ell$,
\begin{align*}
d^{(j)}_{\min}&\triangleq\frac{1}{j}\sum_{i=1}^j\mathbb{E}[(X_i-\mathbb{E}[X_i|S_1,\cdots,S_j])^2]\\
&=\frac{1}{j}d^{(j)}_{\min,1}+\frac{j-1}{j}d_{\min,2},
\end{align*}
where
\begin{align*}
&d^{(j)}_{\min,1}\triangleq\left\{
                      \begin{array}{ll}
                        0, & \lambda^{(j)}_{S,1}=0, \\
                        \frac{\lambda^{(j)}_{X,1}\lambda^{(j)}_{Z,1}}{\lambda^{(j)}_{S,1}}, & \mbox{otherwise},
                      \end{array}
                    \right.\\
&d_{\min,2}\triangleq\left\{
                      \begin{array}{ll}
                        0, & \lambda_{S,2}=0, \\
                        \frac{\lambda_{X,2}\lambda_{Z,2}}{\lambda_{S,2}}, & \mbox{otherwise}.
                      \end{array}
                    \right.
\end{align*}
It is clear that $d_j>d^{(j)}_{\min}$, $j=k,\cdots,\ell$, for any $(r,d_k,\cdots,d_{\ell})\in\mathcal{RD}_k$. Moreover, if $d_j\geq\gamma_X$ for some $j\in\{k,\cdots,\ell\}$, then the corresponding distortion constraint is redundant. Henceforth we shall focus on the case $d_j\in(d^{(j)}_{\min},\gamma_X)$, $j=k,\cdots,\ell$.
\end{remark}

\begin{definition}
For $d_{\ell}\in(d^{(\ell)}_{\min},\gamma_X)$, let
\begin{align*}
r^{(\ell)}(d_{\ell})\triangleq\min\{r: (r,d_{\ell})\in\mathcal{RD}_{\ell}\}.
\end{align*}
\end{definition}

In order to state our main results, we introduce the following quantities.
For any $k\in\{1,\cdots,\ell\}$ and $d_k\in(d^{(k)}_{\min},\gamma_X)$, let
\begin{align*}
&\overline{r}^{(k)}(d_k)\triangleq\frac{1}{2k}\log\frac{(\lambda^{(k)}_{S,1}+\lambda^{(k)}_Q)(\lambda_{S,2}+\lambda^{(k)}_Q)^{k-1}}{(\lambda^{(k)}_Q)^k},\\
&d^{(k)}_j(d_k)\triangleq\frac{\lambda^{(j)}_{X,1}(\lambda^{(j)}_{Z,1}+\lambda^{(k)}_Q)}{j(\lambda^{(j)}_{S,1}+\lambda^{(k)}_Q)}\\
&\hspace{0.5in}+\frac{(j-1)\lambda_{X,2}(\lambda_{Z,2}+\lambda^{(k)}_Q)}{j(\lambda_{S,2}+\lambda^{(k)}_Q)},\quad j=k,\cdots,\ell,
\end{align*}
where $\lambda^{(k)}_Q$ is the unique positive number satisfying
\begin{align}
\frac{\lambda^{(k)}_{X,1}(\lambda^{(k)}_{Z,1}+\lambda^{(k)}_Q)}{k(\lambda^{(k)}_{S,1}+\lambda^{(k)}_Q)}+\frac{(k-1)\lambda_{X,2}(\lambda_{Z,2}+\lambda^{(k)}_Q)}{k(\lambda_{S,2}+\lambda^{(k)}_Q)}=d_k.\label{eq:lambdaQ}
\end{align}

Our first result is a partial characterization of $r^{(\ell)}(d_{\ell})$.
\begin{theorem}\label{thm:theorem1}
For $d_{\ell}\in(d^{(\ell)}_{\min},\gamma_X)$,
\begin{align*}
r^{(\ell)}(d_{\ell})=\overline{r}^{(\ell)}(d_{\ell})
\end{align*}
if either of the following conditions is satisfied:
\begin{enumerate}
\item $\rho_S\geq 0$  and
\begin{align}
&(\ell-1)\lambda^2_{X,2}(\lambda^{(\ell)}_{S,1})^2\mu^{(\ell)}(\mu^{(\ell)}-1)\nonumber\\
&+\ell(\lambda^{(\ell)}_{X,1})^2\lambda^2_{S,2}\geq0, \label{eq:equivcond1l}
\end{align}
where
\begin{align}
\mu^{(\ell)}\triangleq\frac{\lambda_{S,2}-\lambda_{S,2}(\lambda_{S,2}+\lambda^{(\ell)}_Q)^{-1}\lambda_{S,2}}{\lambda^{(\ell)}_{S,1}-\lambda^{(\ell)}_{S,1}(\lambda^{(\ell)}_{S,1}+\lambda^{(\ell)}_Q)^{-1}\lambda^{(\ell)}_{S,1}}.\label{eq:defmu}
\end{align}

\item $\rho_S\leq 0$ and
\begin{align}
(\lambda^{(\ell)}_{X,1})^2\lambda^2_{S,2}\nu^{(\ell)}(\nu^{(\ell)}-1)+\ell\lambda^2_{X,2}(\lambda^{(\ell)}_{S,1})^2\geq0, \label{eq:equivcond2l}
\end{align}
where
\begin{align}
\nu^{(\ell)}\triangleq\frac{\lambda^{(\ell)}_{S,1}-\lambda^{(\ell)}_{S,1}(\lambda^{(\ell)}_{S,1}+\lambda^{(\ell)}_Q)^{-1}\lambda^{(\ell)}_{S,1}}{\lambda_{S,2}-\lambda_{S,2}(\lambda_{S,2}+\lambda^{(\ell)}_Q)^{-1}\lambda_{S,2}}.\label{eq:defnu}
\end{align}
\end{enumerate}
\end{theorem}
\begin{remark}\label{rem:remark3}
\begin{enumerate}
\item Consider the case $\rho_S\geq 0$. When $(\ell-1)\lambda^2_{X,2}(\lambda^{(\ell)}_{S,1})^2\leq 4\ell(\lambda^{(\ell)}_{X,1})^2\lambda^2_{S,2}$, the inequality (\ref{eq:equivcond1l}) always holds, and $r^{(\ell)}(d_{\ell})$ is characterized for all $d_{\ell}\in(d^{(\ell)}_{\min},\gamma_X)$. When $(\ell-1)\lambda^2_{X,2}(\lambda^{(\ell)}_{S,1})^2> 4\ell(\lambda^{(\ell)}_{X,1})^2\lambda^2_{S,2}$, the equation $(\ell-1)\lambda^2_{X,2}(\lambda^{(\ell)}_{S,1})^2\mu^{(\ell)}(\mu^{(\ell)}-1)+\ell(\lambda^{(\ell)}_{X,1})^2\lambda^2_{S,2}=0$ has two real roots in the interval $[0,1]$:
\begin{align*}
&\mu^{(\ell)}_1\triangleq\frac{1}{2}-\frac{1}{2}\sqrt{1-\frac{4\ell(\lambda^{(\ell)}_{X,1})^2\lambda^2_{S,2}}{(\ell-1)\lambda^2_{X,2}(\lambda^{(\ell)}_{S,1})^2}},\\
&\mu^{(\ell)}_2\triangleq\frac{1}{2}+\frac{1}{2}\sqrt{1-\frac{4\ell(\lambda^{(\ell)}_{X,1})^2\lambda^2_{S,2}}{(\ell-1)\lambda^2_{X,2}(\lambda^{(\ell)}_{S,1})^2}}.
\end{align*}
Therefore, the inequality (\ref{eq:equivcond1l}) holds if
\begin{align}
\mu^{(\ell)}\leq\mu^{(\ell)}_1\mbox{ or }\mu^{(\ell)}\geq\mu^{(\ell)}_2.\label{eq:matchingcondition}
\end{align}
It is easy to verify that (\ref{eq:matchingcondition}) is satisfied when $\lambda^{(\ell)}_{S,1}>\lambda_{S,2}=0$ (which implies $\mu^{(\ell)}=0$) or $\lambda^{(\ell)}_{S,1}=\lambda_{S,2}>0$ (which implies $\mu^{(\ell)}=1$). When $\lambda^{(\ell)}_{S,1}>\lambda_{S,2}>0$, $\mu^{(\ell)}$ is a strictly decreasing function of $d_{\ell}$, converging to $1$ as $d_{\ell}\rightarrow d^{(\ell)}_{\min}$ and to $\frac{\lambda_{S,2}}{\lambda^{(\ell)}_{S,1}}$ as $d_{\ell}\rightarrow\gamma_X$; hence, it suffices to analyze the following four scenarios.
\begin{enumerate}
\item $\mu^{(\ell)}_2\leq\frac{\lambda_{S,2}}{\lambda^{(\ell)}_{S,1}}$: $\mu^{(\ell)}\geq\mu^{(\ell)}_2$ is satisfied for all $d_{\ell}\in(d^{(\ell)}_{\min},\gamma_X)$.

\item $\mu^{(\ell)}_1\leq\frac{\lambda_{S,2}}{\lambda^{(\ell)}_{S,1}}$ and $\frac{\lambda_{S,2}}{\lambda^{(\ell)}_{S,1}}<\mu^{(\ell)}_2<1$: $\mu^{(\ell)}\geq\mu^{(\ell)}_2$ is satisfied for all $d_{\ell}$ sufficiently close to $d^{(\ell)}_{\min}$.

\item $\mu^{(\ell)}_1>\frac{\lambda_{S,2}}{\lambda^{(\ell)}_{S,1}}$ and $\mu^{(\ell)}_2<1$: $\mu^{(\ell)}\leq\mu^{(\ell)}_1$ is satisfied for all $d_{\ell}$ sufficiently close to $\gamma_X$ while $\mu^{(\ell)}\geq\mu^{(\ell)}_2$ is satisfied for all $d_{\ell}$ sufficiently close to $d^{(\ell)}_{\min}$.

\item $\mu^{(\ell)}_1=0$ and $\mu^{(\ell)}_2=1$: This can happen only when $\lambda^{(\ell)}_{X,1}=0$.
\end{enumerate}
In view of the above discussion, under the condition $\rho_S\geq 0$, $r^{(\ell)}(d_{\ell})$ is characterized at least for all $d_{\ell}$ sufficiently close to $d^{(\ell)}_{\min}$ unless $\lambda^{(\ell)}_{X,1}=0$ and $\lambda^{(\ell)}_{S,1}>\lambda_{S,2}$ (note that $\lambda^{(\ell)}_{X,1}=0$ implies $\lambda_{S,2}>0$).

\item Consider the case $\rho_S\leq 0$. When $(\lambda^{(\ell)}_{X,1})^2\lambda^2_{S,2}\leq 4\ell\lambda^2_{X,2}(\lambda^{(\ell)}_{S,1})^2$, the inequality (\ref{eq:equivcond2l}) always holds, and $r^{(\ell)}(d_{\ell})$ is characterized for all $d_{\ell}\in(d^{(\ell)}_{\min},\gamma_X)$. When $(\lambda^{(\ell)}_{X,1})^2\lambda^2_{S,2}> 4\ell\lambda^2_{X,2}(\lambda^{(\ell)}_{S,1})^2$, the equation $(\lambda^{(\ell)}_{X,1})^2\lambda^2_{S,2}\nu^{(\ell)}(\nu^{(\ell)}-1)+\ell\lambda^2_{X,2}(\lambda^{(\ell)}_{S,1})^2=0$ has two real roots in the interval $[0,1]$:
\begin{align*}
&\nu^{(\ell)}_1\triangleq\frac{1}{2}-\frac{1}{2}\sqrt{1-\frac{4\ell\lambda^2_{X,2}(\lambda^{(\ell)}_{S,1})^2}{(\lambda^{(\ell)}_{X,1})^2\lambda^2_{S,2}}},\\
&\nu^{(\ell)}_2\triangleq\frac{1}{2}+\frac{1}{2}\sqrt{1-\frac{4\ell\lambda^2_{X,2}(\lambda^{(\ell)}_{S,1})^2}{(\lambda^{(\ell)}_{X,1})^2\lambda^2_{S,2}}}.
\end{align*}
Therefore, the inequality (\ref{eq:equivcond2l})  holds if
\begin{align}
\nu^{(\ell)}\leq\nu^{(\ell)}_1\mbox{ or }\nu^{(\ell)}\geq\nu^{(\ell)}_2.\label{eq:matchingcondition2}
\end{align}
It is easy to verify that (\ref{eq:matchingcondition2}) is satisfied when $\lambda_{S,2}>\lambda^{(\ell)}_{S,1}=0$ (which implies $\nu^{(\ell)}=0$) or $\lambda^{(\ell)}_{S,1}=\lambda_{S,2}>0$ (which implies $\nu^{(\ell)}=1$). When $\lambda_{S,2}>\lambda^{(\ell)}_{S,1}>0$, $\nu^{(\ell)}$ is a strictly decreasing function of $d_{\ell}$, converging to $1$ as $d_{\ell}\rightarrow d^{(\ell)}_{\min}$ and to $\frac{\lambda^{(\ell)}_{S,1}}{\lambda_{S,2}}$ as $d_{\ell}\rightarrow\gamma_X$; hence, it suffices to analyze the following four  scenarios.
\begin{enumerate}
\item $\nu^{(\ell)}_2\leq\frac{\lambda^{(\ell)}_{S,1}}{\lambda_{S,2}}$: $\nu^{(\ell)}\geq\nu^{(\ell)}_2$ is satisfied for all $d_{\ell}\in(d^{(\ell)}_{\min},\gamma_X)$.

\item $\nu^{(\ell)}_1\leq\frac{\lambda^{(\ell)}_{S,1}}{\lambda_{S,2}}$ and $\frac{\lambda^{(\ell)}_{S,1}}{\lambda_{S,2}}<\nu^{(\ell)}_2<1$: $\nu^{(\ell)}\geq\nu^{(\ell)}_2$ is satisfied for all $d_{\ell}$ sufficiently close to $d^{(\ell)}_{\min}$.

\item $\nu^{(\ell)}_1>\frac{\lambda^{(\ell)}_{S,1}}{\lambda_{S,2}}$ and $\nu^{(\ell)}_2<1$: $\nu^{(\ell)}\leq\nu^{(\ell)}_1$ is satisfied for all $d_{\ell}$ sufficiently close to $\gamma_X$ while $\nu^{(\ell)}\geq\nu^{(\ell)}_2$ is satisfied for all $d_{\ell}$ sufficiently close to $d^{(\ell)}_{\min}$.

\item $\nu^{(\ell)}_1=0$ and $\nu^{(\ell)}_2=1$: This can happen only when $\lambda_{X,2}=0$.
\end{enumerate}
In view of the above discussion, under the condition $\rho_S\leq 0$, $r^{(\ell)}(d_{\ell})$ is characterized at least for all $d_{\ell}$ sufficiently close to $d^{(\ell)}_{\min}$ unless $\lambda_{X,2}=0$ and $\lambda_{S,2}>\lambda^{(\ell)}_{S,1}$ (note that $\lambda_{X,2}=0$ implies $\lambda^{(\ell)}_{S,1}>0$).
\end{enumerate}
\end{remark}

Theorem \ref{thm:theorem1}  is a special case of the following more general result.
\begin{theorem}\label{thm:theorem2}
\begin{enumerate}
\item For $d_k\in(d^{(k)}_{\min},\gamma_X)$,
\begin{align*}
(\overline{r}^{(k)}(d_k),d^{(k)}_k(d_k),\cdots,d^{(k)}_{\ell}(d_k))\in\mathcal{RD}_k.
\end{align*}

\item For $(r,d_k,\cdots,d_{\ell})\in\mathcal{RD}_k$ with  $d_k\in(d^{(k)}_{\min},\gamma_X)$,
\begin{align*}
r\geq\overline{r}^{(k)}(d_k)
\end{align*}
if either of the following conditions is satisfied:
\begin{enumerate}
\item[i)] $\rho_S\geq 0$  and
\begin{align}
&(k-1)\lambda^2_{X,2}(\lambda^{(k)}_{S,1})^2\mu^{(k)}(\mu^{(k)}-1)\nonumber\\
&+k(\lambda^{(k)}_{X,1})^2\lambda^2_{S,2}\geq0, \label{eq:equivcond1}
\end{align}
where $\mu^{(k)}$ is defined in (\ref{eq:defmu}) with $\ell$ replaced by $k$.

\item[ii)] $\rho_S\leq 0$ and
\begin{align}
&(\lambda^{(k)}_{X,1})^2\lambda^2_{S,2}\nu^{(k)}(\nu^{(k)}-1)+k\lambda^2_{X,2}(\lambda^{(k)}_{S,1})^2\nonumber\\
&\geq0, \label{eq:equivcond2}
\end{align}
where $\nu^{(k)}$ is defined in (\ref{eq:defnu}) with $\ell$ replaced by $k$.
\end{enumerate}

\item For $j\in\{k,\cdots,\ell\}$ and $(r,d_k,\cdots,d_{\ell})\in\mathcal{RD}_k$ with  $d_k\in(d^{(k)}_{\min},\gamma_X)$ and $r=\overline{r}^{(k)}(d_k)$, we have
\begin{align*}
d_j\geq d^{(k)}_j(d_k)
\end{align*}
if either of the following conditions is satisfied:
\begin{enumerate}
\item Condition i).

\item $\rho_S\leq 0$, $\lambda^{(j)}_{S,1}>0$, and
\begin{align}
&(\nu^{(k,j)}+(k-1))(\lambda^{(k)}_{X,1})^2\lambda^2_{S,2}(\nu^{(k)})^2\nonumber\\
&+(k-1)(\nu^{(k,j)}-\nu^{(k)})\lambda^2_{X,2}(\lambda^{(k)}_{S,1})^2\geq0,\label{eq:equivcond3}\\
&(\nu^{(k,j)}-1)(\lambda^{(k)}_{X,1})^2\lambda^2_{S,2}(\nu^{(k)})^2\nonumber\\
&+((k-1)\nu^{(k,j)}+\nu^{(k)})\lambda^2_{X,2}(\lambda^{(k)}_{S,1})^2\geq0,\label{eq:equivcond4}
\end{align}
where
\begin{align*}
\nu^{(k,j)}=\frac{\lambda^{(j)}_{S,1}-\lambda^{(j)}_{S,1}(\lambda^{(j)}_{S,1}+\lambda^{(k)}_Q)^{-1}\lambda^{(j)}_{S,1}}{\lambda_{S,2}-\lambda_{S,2}(\lambda_{S,2}+\lambda^{(k)}_Q)^{-1}\lambda_{S,2}}.
\end{align*}
\end{enumerate}
\end{enumerate}
\end{theorem}
\begin{IEEEproof}
See Section \ref{sec:theorem2}.
\end{IEEEproof}

\begin{remark}
\begin{enumerate}
\item The argument in Remark \ref{rem:remark3} can be leveraged to prove that, for the case $\rho_S\geq 0$, the inequality (\ref{eq:equivcond1}) holds at least for all $d_{k}$ sufficiently close to $d^{(k)}_{\min}$ unless $\lambda^{(k)}_{X,1}=0$ (which can happen only when $k=\ell$) and $\lambda^{(k)}_{S,1}>\lambda_{S,2}$ (note that $\lambda^{(k)}_{X,1}=0$ implies $\lambda_{S,2}>0$); similarly, for the case $\rho_S\leq 0$, the inequality (\ref{eq:equivcond2}) holds at least for all $d_{k}$ sufficiently close to $d^{(k)}_{\min}$ unless $\lambda_{X,2}=0$ and $\lambda_{S,2}>\lambda^{(k)}_{S,1}$ (note that $\lambda_{X,2}=0$ implies $\lambda^{(k)}_{S,1}>0$).

\item For the case $\rho_S\leq 0$, the condition $\lambda^{(j)}_{S,1}>0$ can be potentially violated (i.e., $\lambda^{(j)}_{S,1}=0$) only when $j=\ell$.

\item Consider the case $\rho_S\leq 0$ and $\lambda^{(j)}_{S,1}>0$. If $\lambda^{(k)}_{X,1}>0$, then the inequality (\ref{eq:equivcond3}) holds at least for $d_{k}$ sufficiently close to $d^{(k)}_{\min}$; if  $\lambda^{(k)}_{X,1}=0$, which implies $k=j=\ell$, then the inequality (\ref{eq:equivcond3}) always holds. The inequality (\ref{eq:equivcond4}) holds at least for $d_{k}$ sufficiently close to $d^{(k)}_{\min}$ unless $\lambda_{X,2}=0$ and $\lambda_{S,2}>\lambda^{(j)}_{S,1}$.

\end{enumerate}
\end{remark}

\section{Proof of Theorem \ref{thm:theorem2}}\label{sec:theorem2}

The following lemma can be obtained by adapting the classical result by Berger \cite{Berger78} and Tung \cite{Tung78} to the current setting.

\begin{lemma}\label{lem:robustBergerTung}
For any auxiliary random vector $V\triangleq(V_1,\cdots,V_{\ell})^T$ jointly distributed with $(X,Z,S)$ such that $\{X,Z,(S_{i'})_{i'\in\{1,\cdots,\ell\}\backslash\{i\}},(V_{i'})_{i'\in\{1,\cdots,\ell\}\backslash\{i\}}\}\leftrightarrow S_i\leftrightarrow V_i$ form a Markov chain, $i=1,\cdots,\ell$, and any $(r,d_k\cdots,d_{\ell})$ such that
\begin{align*}
&r1_k\in\mathcal{R}(\mathcal{A}),\quad\mathcal{A}\subseteq\{1,\cdots,\ell\}\mbox{ with }|\mathcal{A}|=k,\\
&d_{|\mathcal{A}|}\geq\frac{1}{|\mathcal{A}|}\sum\limits_{i\in\mathcal{A}}\mathbb{E}[(X_i-\mathbb{E}[X_i|(V_{i'})_{i'\in\mathcal{A}}])^2],\\
&\hspace{1in}\mathcal{A}\subseteq\{1,\cdots,\ell\}\mbox{ with }|\mathcal{A}|\geq k,
\end{align*}
where $\mathcal{R}(\mathcal{A})$ denotes the set of $(r_i)_{i\in\mathcal{A}}$ satisfying
\begin{align*}
&\sum\limits_{i\in\mathcal{B}}r_i\geq I((S_i)_{i\in\mathcal{B}};(V_i)_{i\in\mathcal{B}}|(V_{i})_{i\in\mathcal{A}\backslash\mathcal{B}}),\quad \emptyset\subset\mathcal{B}\subseteq\mathcal{A},
\end{align*}
we have
\begin{align*}
(r,d_k\cdots,d_{\ell})\in\mathcal{RD}_k.
\end{align*}
\end{lemma}

Equipped with this lemma, we are in a position to prove Part 1) of Theorem \ref{thm:theorem2}. Let $Q\triangleq(Q_1,\cdots,Q_{\ell})^T$ be an $\ell$-dimensional zero-mean Gaussian random vector with covariance matrix
\begin{align*}
\Lambda_Q\triangleq\mathrm{diag}^{(\ell)}(\lambda_Q,\cdots,\lambda_Q)\succ 0.
\end{align*}
Moreover, we assume that $Q$ is independent of $(X,Z,S)$, and let
\begin{align*}
V_i\triangleq S_i+Q_i,\quad i=1,\cdots,\ell.
\end{align*}
Clearly, $V\triangleq(V_1,\cdots,V_{\ell})^T$ satisfies the condition specified in Lemma \ref{lem:robustBergerTung}. Let
\begin{align*}
&r\triangleq\frac{1}{k}I(S_1,\cdots,S_k;V_1,\cdots,V_k),\\
&d_j\triangleq\frac{1}{j}\sum\limits_{i=1}^j\mathbb{E}[(X_i-\mathbb{E}[X_i|V_1,\cdots,V_j])^2],\\
&\hspace{1.7in}  j=k,\cdots,\ell.
\end{align*}
It is easy to show that $r1_k\in\mathcal{R}(\mathcal{A})$ for all $\mathcal{A}\subseteq\{1,\cdots,\ell\}$ with $|\mathcal{A}|=k$ by leveraging the contra-polymatroid structure \cite{XCWB07} of $\mathcal{R}(\mathcal{A})$ and the symmetry of the underlying distributions. Let $\Lambda^{(j)}_Q$ denote the leading $j\times j$ principal submatrix of $\Lambda_Q$, $j=k,\cdots,\ell$. We have
\begin{align*}
r&=\frac{1}{k}(h(V_1,\cdots,V_k)-h(V_1,\cdots,V_k|S_1,\cdots,S_k))\\
&=\frac{1}{k}(h(S_1+Q_1,\cdots,S_k+Q_k)-h(Q_1,\cdots,Q_k))\\
&=\frac{1}{2k}\log\frac{\det(\Gamma^{(k)}_S+\Lambda^{(k)}_Q)}{\det(\Lambda^{(k)}_Q)}\\
&=\frac{1}{2k}\log\frac{\det(\Lambda^{(k)}_S+\Lambda^{(k)}_Q)}{\det(\Lambda^{(k)}_Q)}\\
&=\frac{1}{2k}\log\frac{(\lambda^{(k)}_{S,1}+\lambda_Q)(\lambda_{S,2}+\lambda_Q)^{k-1}}{\lambda^{k}_Q}.
\end{align*}
Moreover, for $j=k,\cdots,\ell$,
\begin{align*}
d_j&=\frac{1}{j}\mathrm{tr}(\Gamma^{(j)}_X-\Gamma^{(j)}_X(\Gamma^{(j)}_S+\Lambda^{(j)}_Q)^{-1}\Gamma^{(j)}_X)\\
&=\frac{1}{j}\mathrm{tr}(\Lambda^{(j)}_X-\Lambda^{(j)}_X(\Lambda^{(j)}_S+\Lambda^{(j)}_Q)^{-1}\Lambda^{(j)}_X)\\
&=\frac{\lambda^{(j)}_{X,1}(\lambda^{(j)}_{Z,1}+\lambda_Q)}{j(\lambda^{(j)}_{S,1}+\lambda_Q)}+\frac{(j-1)\lambda_{X,2}(\lambda_{Z,2}+\lambda_Q)}{j(\lambda_{S,2}+\lambda_Q)},
\end{align*}
which is a strictly increasing function of $\lambda_Q$, converging to $d^{(j)}_{\min}$ as $\lambda_Q\rightarrow 0$ and to $\gamma_X$ as $\lambda_Q\rightarrow\infty$. One can readily complete the proof of Part 1) of Theorem \ref{thm:theorem2} by invoking Lemma \ref{lem:robustBergerTung}.

Now we proceed to prove Part 2) and Part 3) of Theorem \ref{thm:theorem2}. Fix $k$ and $j$ with $1\leq k\leq j\leq\ell$.
First consider the case $\Gamma^{(j)}_S\succ0$ (i.e., $\lambda^{(j)}_{S,1}>0$ and $\lambda_{S,2}>0$). Let
$(S_1,\cdots,S_j)^T=(U_1,\cdots,U_j)^T+(W_1,\cdots,W_j)^T$
be a fictitious signal-noise decomposition of $(S_1,\cdots,S_j)^T$, where $(U_1,\cdots,U_j)^T$ and $(W_1,\cdots,W_j)^T$ are two mutually independent $j$-dimensional zero-mean Gaussian vectors with covariance matrices
\begin{align*}
&\Gamma^{(j)}_U\succ0,\\
&\Lambda^{(j)}_W\triangleq\mathrm{diag}^{(j)}(\lambda_W,\cdots,\lambda_W)\succ 0,
\end{align*}
respectively. We then construct the auxiliary random processes $\{(U_1(t),\cdots,U_{j}(t))^T\}_{t=1}^\infty$
and $\{(W_1(t),\cdots,W_{\ell}(t))^T\}_{t=1}^\infty$ accordingly.


It is worth mentioning that the idea of augmenting the probability space via the introduction of auxiliary random processes is inspired by  \cite{WTV08, WCW10, WC13, WC14, Oohama14, Ozarow80, WV07,WV09,Chen09}.
Our construction (without the symmetry constraint) can be viewed as a generalization of that in \cite{WCW10}, which is restricted to the special case where the corruptive noises are absent. It should also be contrasted with the conventional approach where $(U_1,\cdots,U_j)^T$ and $(W_1,\cdots,W_j)^T$ are set respectively to be $(X_1,\cdots,X_j)^T$ and $(Z_1,\cdots,Z_j)^T$ (with the components of $(Z_1,\cdots,Z_j)^T$ assumed to be mutually independent); our construction is more flexible and often yields stronger results. The fictitious signal-noise decomposition is closely related to the Markov coupling argument in \cite{WA08}.
One subtle difference is that the fictitious decomposition is specified for $(S_1,\cdots,S_j)^T$ instead of $(S_1,\cdots,S_{\ell})^T$. As a consequence, we can choose $\lambda_W$ from $(0,\min\{\lambda^{(j)}_{S,1},\lambda_{S,2}\})$, which may offer more freedom than $(0,\min\{\lambda^{(\ell)}_{S,1},\lambda_{S,2}\})$ since $\min\{\lambda^{(j)}_{S,1},\lambda_{S,2}\}\geq\min\{\lambda^{(\ell)}_{S,1},\lambda_{S,2}\}$ and the inequality is strict when $\rho_S<0$ and $j<\ell$.

In view of Definition \ref{def:robust}, for any $(r,d_k\cdots,d_{\ell})\in\mathcal{RD}_k$ and $\epsilon>0$,
there exist encoding functions $\phi^{(n)}_i:\mathbb{R}^n\rightarrow\mathcal{C}^{(n)}_i$, $i=1,\cdots,j$, such that
\begin{align}
&\frac{1}{k n}\sum\limits_{i\in\mathcal{A}}\log|\mathcal{C}^{(n)}_i|\leq r+\epsilon,\nonumber\\
&\hspace{1.0in}\mathcal{A}\subseteq\{1,\cdots,j\}\mbox{ with }|\mathcal{A}|=k,\label{eq:applyconstraint1}\\
&\frac{1}{kn}\sum\limits_{i\in\mathcal{A}}\sum\limits_{t=1}^n\mathbb{E}[(X_i(t)-\hat{X}_{i,\mathcal{A}}(t))^2]\leq d_{k}+\epsilon,\nonumber\\
&\hspace{1.0in}\mathcal{A}\subseteq\{1,\cdots,j\}\mbox{ with }|\mathcal{A}|=k,\label{eq:applyconstraint2}\\
&\frac{1}{jn}\sum\limits_{i=1}^j\sum\limits_{t=1}^n\mathbb{E}[(X_i(t)-\hat{X}_{i,\{1,\cdots,j\}}(t))^2]\leq d_{j}+\epsilon.\nonumber
\end{align}
We have
\begin{align}
&\sum\limits_{i\in\mathcal{A}}\log|\mathcal{C}^{(n)}_i|\nonumber\\
&\geq H((\phi^{(n)}_i(S^n_i))_{i\in\mathcal{A}})\nonumber\\
&=I((U^n_i)_{i\in\mathcal{A}};(\phi^{(n)}_i(S^n_i))_{i\in\mathcal{A}})\nonumber\\
&\quad+H((\phi^{(n)}_i(S^n_i))_{i\in\mathcal{A}}|(U^n_i)_{i\in\mathcal{A}})\nonumber\\
&=I((U^n_i)_{i\in\mathcal{A}};(\phi^{(n)}_i(S^n_i))_{i\in\mathcal{A}})\nonumber\\
&\quad+I((S^n_i)_{i\in\mathcal{A}};(\phi^{(n)}_i(S^n_i))_{i\in\mathcal{A}}|(U^n_i)_{i\in\mathcal{A}})\nonumber\\
&=h((U^n_i)_{i\in\mathcal{A}})+h((W^n_i)_{i\in\mathcal{A}})\nonumber\\
&\quad-h((U^n_i)_{i\in\mathcal{A}}|(\phi^{(n)}_i(S^n_i))_{i\in\mathcal{A}})\nonumber\\
&\quad-h((S^n_i)_{i\in\mathcal{A}}|(U^n_i)_{i\in\mathcal{A}},(\phi^{(n)}_i(S^n_i))_{i\in\mathcal{A}})\nonumber\\
&=\frac{n}{2}\log((2\pi e)^k\det(\Gamma^{(k)}_U))+\frac{n}{2}\log((2\pi e)^k\det(\Lambda^{(k)}_W))\nonumber\\
&\quad-h((U^n_i)_{i\in\mathcal{A}}|(\phi^{(n)}_i(S^n_i))_{i\in\mathcal{A}})\nonumber\\
&\quad-h((S^n_i)_{i\in\mathcal{A}}|(U^n_i)_{i\in\mathcal{A}},(\phi^{(n)}_i(S^n_i))_{i\in\mathcal{A}}),\label{eq:tobesub}
\end{align}
where $\Gamma^{(k)}_U$ and $\Lambda^{(k)}_W$ denote the leading $k\times k$ principal submatrices of $\Gamma^{(j)}_U$ and $\Lambda^{(j)}_W$, respectively.
For $t=1,\cdots,n$, let
\begin{align*}
&\Sigma_{\mathcal{A}}(t)\triangleq\mathbb{E}[(U_i(t)-\hat{U}_{i,\mathcal{A}}(t))^T_{i\in\mathcal{A}}(U_i(t)-\hat{U}_{i,\mathcal{A}}(t))_{i\in\mathcal{A}}],\\
&\Delta_{\mathcal{A}}(t)\triangleq\mathbb{E}[(S_i(t)-\tilde{S}_{i,\mathcal{A}}(t))^T_{i\in\mathcal{A}}(S_i(t)-\tilde{S}_{i,\mathcal{A}}(t))_{i\in\mathcal{A}}],
\end{align*}
where
\begin{align*}
&\hat{U}_{i,\mathcal{A}}(t)\triangleq\mathbb{E}[U_{i}(t)|(\phi^{(n)}_{i'}(S^n_{i'}))_{i'\in\mathcal{A}}],\quad i\in\mathcal{A},\\
&\tilde{S}_{i,\mathcal{A}}(t)\triangleq\mathbb{E}[S_i(t)|(U^n_{i'})_{i'\in\mathcal{A}},(\phi^{(n)}_{i'}(S^n_{i'}))_{i'\in\mathcal{A}}],\quad i\in\mathcal{A}.
\end{align*}
Moreover, let
\begin{align*}
&\Sigma_{\mathcal{A}}\triangleq\frac{1}{n}\sum\limits_{t=1}^n\Sigma_{\mathcal{A}}(t),\\
&\Delta_{\mathcal{A}}\triangleq\frac{1}{n}\sum\limits_{t=1}^n\Delta_{\mathcal{A}}(t).
\end{align*}
It can be verified that
\begin{align}
&h((U^n_i)_{i\in\mathcal{A}}|(\phi^{(n)}_i(S^n_i))_{i\in\mathcal{A}})\nonumber\\
&=\sum\limits_{t=1}^nh((U_i(t))_{i\in\mathcal{A}}|(\phi^{(n)}_i(S^n_i))_{i\in\mathcal{A}},(U^{t-1}_i)_{i\in\mathcal{A}})\nonumber\\
&\leq\sum\limits_{t=1}^nh((U_i(t))_{i\in\mathcal{A}}|(\phi^{(n)}_i(S^n_i))_{i\in\mathcal{A}})\nonumber\\
&=\sum\limits_{t=1}^nh((U_i(t)-\hat{U}_{i,\mathcal{A}}(t))_{i\in\mathcal{A}}|(\phi^{(n)}_i(S^n_i))_{i\in\mathcal{A}})\nonumber\\
&\leq\sum\limits_{t=1}^nh((U_i(t)-\hat{U}_{i,\mathcal{A}}(t))_{i\in\mathcal{A}})\nonumber\\
&\leq\sum\limits_{t=1}^n\frac{1}{2}\log((2\pi e)^k\det(\Sigma_{\mathcal{A}}(t)))\label{eq:maxentropy}\\
&\leq\frac{n}{2}\log((2\pi e)^k\det(\Sigma_{\mathcal{A}})),\label{eq:concavitylog}
\end{align}
where (\ref{eq:maxentropy}) is due to the maximum differential entropy lemma \cite[p. 21]{EGK11}, and (\ref{eq:concavitylog}) is due to the concavity of the  log-determinant function. Similarly, we have
\begin{align}
&h((S^n_i)_{i\in\mathcal{A}}|(U^n_i)_{i\in\mathcal{A}},(\phi^{(n)}_i(S^n_i))_{i\in\mathcal{A}})\nonumber\\
&\leq\frac{n}{2}\log((2\pi e)^k\det(\Delta_{\mathcal{A}})).\label{eq:similarderivation}
\end{align}
Combining (\ref{eq:applyconstraint1}), (\ref{eq:tobesub}), (\ref{eq:concavitylog}), and (\ref{eq:similarderivation}) gives
\begin{align}
\frac{1}{2k}\log\frac{\det(\Gamma^{(k)}_U)\det(\Lambda^{(k)}_W)}{\det(\Sigma_{\mathcal{A}})\det(\Delta_{\mathcal{A}})}\leq r+\epsilon.\label{eq:rconstraint}
\end{align}
For $t=1,\cdots,n$, let
\begin{align*}
D_{\mathcal{A}}(t)\triangleq\mathbb{E}[(S_i(t)-\hat{S}_{i,\mathcal{A}}(t))^T_{i\in\mathcal{A}}(S_i(t)-\hat{S}_{i,\mathcal{A}}(t))_{i\in\mathcal{A}}],
\end{align*}
where
\begin{align*}
&\hat{S}_{i,\mathcal{A}}(t)\triangleq\mathbb{E}[S_i(t)|(\phi^{(n)}_{i'}(S^n_{i'}))_{i'\in\mathcal{A}}],\quad i\in\mathcal{A}.
\end{align*}
Moreover, let
\begin{align*}
D_{\mathcal{A}}\triangleq\frac{1}{n}\sum\limits_{t=1}^nD_{\mathcal{A}}(t).
\end{align*}
Clearly, we have
\begin{align}
0\prec D_{\mathcal{A}}\preceq\Gamma^{(k)}_S.\label{eq:trivial1}
\end{align}
Furthermore, as shown in Appendix \ref{app:constraints},
\begin{align}
\Sigma_{\mathcal{A}}&=\Gamma^{(k)}_U(\Gamma^{(k)}_S)^{-1}D_{\mathcal{A}}(\Gamma^{(k)}_S)^{-1}\Gamma^{(k)}_U+\Gamma^{(k)}_U\nonumber\\
&\quad-\Gamma^{(k)}_U(\Gamma^{(k)}_S)^{-1}\Gamma^{(k)}_U,\label{eq:sigma}\\
\Delta_{\mathcal{A}}&\preceq(D^{-1}_{\mathcal{A}}+(\Lambda^{(k)}_W)^{-1}-(\Gamma^{(k)}_S)^{-1})^{-1}.\label{eq:delta1}
\end{align}
The argument for (\ref{eq:sigma}) can also be leveraged to prove
\begin{align*}
&\frac{1}{n}\sum\limits_{i\in\mathcal{A}}\sum\limits_{t=1}^n\mathbb{E}[(X_i(t)-\hat{X}_{i,\mathcal{A}}(t))^2]\\
&=\mathrm{tr}(\Gamma^{(k)}_X(\Gamma^{(k)}_S)^{-1}D_{\mathcal{A}}(\Gamma^{(k)}_S)^{-1}\Gamma^{(k)}_X+\Gamma^{(k)}_X\\
&\quad-\Gamma^{(k)}_X(\Gamma^{(k)}_S)^{-1}\Gamma^{(k)}_X),
\end{align*}
which, together with (\ref{eq:applyconstraint2}), implies
\begin{align}
&\mathrm{tr}(\Gamma^{(k)}_X(\Gamma^{(k)}_S)^{-1}D_{\mathcal{A}}(\Gamma^{(k)}_S)^{-1}\Gamma^{(k)}_X+\Gamma^{(k)}_X\nonumber\\
&-\Gamma^{(k)}_X(\Gamma^{(k)}_S)^{-1}\Gamma^{(k)}_X)\leq k(d_k+\epsilon).\label{eq:dconstraint}
\end{align}
For $t=1,\cdots,n$, let
\begin{align*}
&\Delta_{\{1,\cdots,j\}}(t)\\
&\triangleq\mathbb{E}[(S_1(t)-\tilde{S}_{1,\{1,\cdots,j\}}(t),\cdots,S_j(t)-\tilde{S}_{j,\{1,\cdots,j\}}(t))^T\\
&\hspace{0.34in}(S_1(t)-\tilde{S}_{1,\{1,\cdots,j\}}(t),\cdots,S_j(t)-\tilde{S}_{j,\{1,\cdots,j\}}(t))],\\
&D_{\{1,\cdots,j\}}(t)\\
&\triangleq\mathbb{E}[(S_1(t)-\hat{S}_{1,\{1,\cdots,j\}}(t),\cdots,S_j(t)-\hat{S}_{j,\{1,\cdots,j\}}(t))^T\\
&\hspace{0.34in}(S_1(t)-\hat{S}_{1,\{1,\cdots,j\}}(t),\cdots,S_j(t)-\hat{S}_{j,\{1,\cdots,j\}}(t))],\\
&\delta_i(t)\triangleq\mathbb{E}[(S_i(t)-\tilde{S}_i(t))^2],\quad i=1,\cdots,j,
\end{align*}
where
\begin{align*}
&\tilde{S}_{i,\{1,\cdots,j\}}(t)\\
&\triangleq\mathbb{E}[S_i(t)|U^n_1,\cdots,U^n_j,\phi^{(n)}_1(S^n_1),\cdots,\phi^{(n)}_j(S^n_j)],\\
&\hspace{2.1in} i=1,\cdots,j,\\
&\hat{S}_{i,\{1,\cdots,j\}}(t)\triangleq\mathbb{E}[S_i(t)|\phi^{(n)}_1(S^n_1),\cdots,\phi^{(n)}_j(S^n_j)],\nonumber\\
&\hspace{2.1in} i=1,\cdots,j,\\
&\tilde{S}_i(t)\triangleq\mathbb{E}[S_i(t)|U^n_i,\phi^{(n)}_i(S^n_i)],\quad i=1,\cdots,j.
\end{align*}
Moreover, let
\begin{align*}
&\Delta_{\{1,\cdots,j\}}\triangleq\frac{1}{n}\sum\limits_{t=1}^n\Delta_{\{1,\cdots,j\}}(t),\\
&D_{\{1,\cdots,j\}}\triangleq\frac{1}{n}\sum\limits_{t=1}^nD_{\{1,\cdots,j\}}(t),\\
&\delta_i\triangleq\sum\limits_{t=1}^n\delta_i(t),\quad i=1,\cdots,j.
\end{align*}
The argument for (\ref{eq:delta1}) and (\ref{eq:dconstraint}) can be leveraged to show that
\begin{align}
&\Delta_{\{1,\cdots,j\}}\preceq(D^{-1}_{\{1,\cdots,j\}}+(\Lambda^{(j)}_W)^{-1}-(\Gamma^{(j)}_S)^{-1})^{-1},\label{eq:delta1j}\\
&\mathrm{tr}(\Gamma^{(j)}_X(\Gamma^{(j)}_S)^{-1}D_{\{1,\cdots,j\}}(\Gamma^{(j)}_S)^{-1}\Gamma^{(j)}_X+\Gamma^{(j)}_X\nonumber\\
&-\Gamma^{(j)}_X(\Gamma^{(j)}_S)^{-1}\Gamma^{(j)}_X)\leq j(d_j+\epsilon).\label{eq:dconstraintj}
\end{align}
It is also clear that
\begin{align}
0<\delta_i,\quad i=1,\cdots,\ell.\label{eq:trivial2}
\end{align}
Furthermore, in view of the fact that $S^n_i=U^n_i+W^n_i$, $i=1,\cdots,j$, and that $(U^n_1,\cdots,U^n_j), W^n_1,\cdots,W^n_{j}$ are mutually independent, we must have
\begin{align}
&\Delta_{\mathcal{A}}=\mathrm{diag}^{(k)}(\delta_i)_{i\in\mathcal{A}},\label{eq:delta2A}\\
&\Delta_{\{1,\cdots,j\}}=\mathrm{diag}^{(j)}(\delta_1,\cdots,\delta_j).\label{eq:delta2}
\end{align}
Combining (\ref{eq:rconstraint})--(\ref{eq:delta2}), sending $\epsilon\rightarrow 0$, and invoking a symmetrization and convexity argument shows that there exist $D^{(k)}$, $D^{(j)}$, and $\delta$ satisfying the following set of inequalities
\begin{align}
&\frac{1}{2k}\log\frac{\det(\Gamma^{(k)}_U)}{\det(\Sigma^{(k)})}+\frac{1}{2}\log\frac{\lambda_W}{\delta}\leq r,\label{eq:new1}\\
&0\prec D^{(k)}\preceq\Gamma^{(k)}_S,\label{eq:new2}\\
&0<\delta,\label{eq:new3}\\
&\mathrm{diag}^{(k)}(\delta,\cdots,\delta)\nonumber\\
&\preceq((D^{(k)})^{-1}+(\Lambda^{(k)}_W)^{-1}-(\Gamma^{(k)}_S)^{-1})^{-1},\label{eq:new4}\\
&\mathrm{tr}(\Gamma^{(k)}_X(\Gamma^{(k)}_S)^{-1}D^{(k)}(\Gamma^{(k)}_S)^{-1}\Gamma^{(k)}_X+\Gamma^{(k)}_X\nonumber\\
&-\Gamma^{(k)}_X(\Gamma^{(k)}_S)^{-1}\Gamma^{(k)}_X)\leq kd_k,\label{eq:new5}\\
&\mathrm{diag}^{(j)}(\delta,\cdots,\delta)\nonumber\\
&\preceq((D^{(j)})^{-1}+(\Lambda^{(j)}_W)^{-1}-(\Gamma^{(j)}_S)^{-1})^{-1},\label{eq:new6}\\
&\mathrm{tr}(\Gamma^{(j)}_X(\Gamma^{(j)}_S)^{-1}D^{(j)}(\Gamma^{(j)}_S)^{-1}\Gamma^{(j)}_X+\Gamma^{(j)}_X\nonumber\\
&-\Gamma^{(j)}_X(\Gamma^{(j)}_S)^{-1}\Gamma^{(j)}_X)\leq jd_j,\label{eq:new7}
\end{align}
where
\begin{align*}
&D^{(k)}=\Theta^{(k)}\mathrm{diag}^{(k)}(d^{(k)}_1,d^{(k)}_2,\cdots,d^{(k)}_2)(\Theta^{(k)})^T,\\
&D^{(j)}=\Theta^{(j)}\mathrm{diag}^{(j)}(d^{(j)}_1,d^{(j)}_2,\cdots,d^{(j)}_2)(\Theta^{(j)})^T
\end{align*}
for some $d^{(k)}_1,d^{(k)}_2,d^{(k)}_1,d^{(k)}_2$, and
\begin{align*}
\Sigma^{(k)}&\triangleq\Gamma^{(k)}_U(\Gamma^{(k)}_S)^{-1}D^{(k)}(\Gamma^{(k)}_S)^{-1}\Gamma^{(k)}_U+\Gamma^{(k)}_U\nonumber\\
&\quad-\Gamma^{(k)}_U(\Gamma^{(k)}_S)^{-1}\Gamma^{(k)}_U.
\end{align*}
Equivalently, (\ref{eq:new1})--(\ref{eq:new7}) can be written as
\begin{align}
&\frac{1}{2k}\log\frac{(\lambda^{(k)}_{S,1})^2}{(\lambda^{(k)}_{S,1}-\lambda_W)d^{(k)}_1+\lambda^{(k)}_{S,1}\lambda_W}\nonumber\\
&+\frac{k-1}{2k}\log\frac{\lambda^2_{S,2}}{(\lambda_{S,2}-\lambda_W)d^{(k)}_2+\lambda_{S,2}\lambda_W}+\frac{1}{2}\log\frac{\lambda_W}{\delta}\nonumber\\
&\leq r,\label{eq:equiv1}\\
&0<d^{(k)}_1\leq\lambda^{(k)}_{S,1},\label{eq:equiv2}\\
&0<d^{(k)}_2\leq\lambda_{S,2},\label{eq:equiv3}\\
&0<\delta,\label{eq:equiv4}\\
&\delta\leq((d^{(k)}_1)^{-1}+\lambda^{-1}_W-(\lambda^{(k)}_{S,1})^{-1})^{-1},\label{eq:equiv5}\\
&\delta\leq((d^{(k)}_2)^{-1}+\lambda^{-1}_W-\lambda^{-1}_{S,2})^{-1},\label{eq:equiv6}\\
&(\lambda^{(k)}_{X,1})^{2}(\lambda^{(k)}_{S,1})^{-2}d^{(k)}_1+\lambda^{(k)}_{X,1}-(\lambda^{(k)}_{X,1})^2(\lambda^{(k)}_{S,1})^{-1}\nonumber\\
&+(k-1)(\lambda^2_{X,2}\lambda^{-2}_{S,2}d^{(k)}_2+\lambda_{X,2}-\lambda^2_{X,2}\lambda^{-1}_{S,2})\nonumber\\
&\leq k d_k,\label{eq:equiv7}\\
&\delta\leq((d^{(j)}_1)^{-1}+\lambda^{-1}_W-(\lambda^{(j)}_{S,1})^{-1})^{-1},\label{eq:equiv8}\\
&\delta\leq((d^{(j)}_2)^{-1}+\lambda^{-1}_W-\lambda^{-1}_{S,2})^{-1},\label{eq:equiv9}\\
&(\lambda^{(j)}_{X,1})^{2}(\lambda^{(j)}_{S,1})^{-2}d^{(j)}_1+\lambda^{(j)}_{X,1}-(\lambda^{(j)}_{X,1})^2(\lambda^{(j)}_{S,1})^{-1}\nonumber\\
&+(j-1)(\lambda^2_{X,2}\lambda^{-2}_{S,2}d^{(j)}_2+\lambda_{X,2}-\lambda^2_{X,2}\lambda^{-1}_{S,2})\nonumber\\
&\leq j d_j.\label{eq:equiv10}
\end{align}

When $\lambda^{(j)}_{S,1}\geq\lambda_{S,2}>0$, we can send $\lambda_W\rightarrow\lambda_{S,2}$  and deduce from (\ref{eq:equiv1}), (\ref{eq:equiv5}), (\ref{eq:equiv6}), (\ref{eq:equiv8}), and (\ref{eq:equiv9}) that
\begin{align}
&\eta(d^{(k)}_1,d^{(k)}_2,\delta)\leq r,\label{eq:1case1}\\
&\delta\leq((d^{(k)}_1)^{-1}+\lambda^{-1}_{S,2}-(\lambda^{(k)}_{S,1})^{-1})^{-1},\label{eq:2case1}\\
&\delta\leq d^{(k)}_2,\label{eq:3case1}\\
&\delta\leq((d^{(j)}_1)^{-1}+\lambda^{-1}_{S,2}-(\lambda^{(j)}_{S,1})^{-1})^{-1},\label{eq:4case1}\\
&\delta\leq d^{(j)}_2,\label{eq:5case1}
\end{align}
where
\begin{align*}
&\eta(d^{(k)}_1,d^{(k)}_2,\delta)\\
&\triangleq\frac{1}{2k}\log\frac{(\lambda^{(k)}_{S,1})^2}{(\lambda^{(k)}_{S,1}-\lambda_{S,2})d^{(k)}_1+\lambda^{(k)}_{S,1}\lambda_{S,2}}+\frac{1}{2}\log\frac{\lambda_{S,2}}{\delta}.
\end{align*}
Furthermore, combining (\ref{eq:equiv10}), (\ref{eq:4case1}), and (\ref{eq:5case1}) gives
\begin{align}
d_j&\geq\frac{1}{j}((\lambda^{(j)}_{X,1})^{2}(\lambda^{(j)}_{S,1})^{-2}(\delta^{-1}+(\lambda^{(j)}_{S,1})^{-1}-\lambda^{-1}_{S,2})^{-1}\nonumber\\
&\qquad+\lambda^{(j)}_{X,1}-(\lambda^{(j)}_{X,1})^2(\lambda^{(j)}_{S,1})^{-1})\nonumber\\
&\quad+\frac{j-1}{j}(\lambda^2_{X,2}\lambda^{-2}_{S,2}\delta+\lambda_{X,2}-\lambda^2_{X,2}\lambda^{-1}_{S,2}).\label{eq:invoke1}
\end{align}
Now consider the following convex optimization problem:
\begin{align*}
\min\limits_{d^{(k)}_1,d^{(k)}_2,\delta}\eta(d^{(k)}_1,d^{(k)}_2,\delta)\qquad\qquad\mbox{(} \mathbf{P} \mbox{)}
\end{align*}
subject to (\ref{eq:equiv2}), (\ref{eq:equiv3}), (\ref{eq:equiv4}), (\ref{eq:2case1}), (\ref{eq:3case1}), and (\ref{eq:equiv7}).
According to the Karush-Kuhn-Tucker conditions, $(d^{(k)}_{1},d^{(k)}_2,\delta)$ is a minimizer of the convex optimization problem ($\mathbf{P}$) if and only if (\ref{eq:equiv2}), (\ref{eq:equiv3}), (\ref{eq:equiv4}), (\ref{eq:2case1}), (\ref{eq:3case1}), and (\ref{eq:equiv7}) are satisfied, and there exist nonnegative $a_1,a_2,b_1,b_2,c$ such that
\begin{align}
&\frac{\lambda_{S,2}-\lambda^{(k)}_{S,1}}{2k ((\lambda^{(k)}_{S,1}-\lambda^{(k)}_{S,2})d^{(k)}_1+\lambda^{(k)}_{S,1}\lambda_{S,2})}+a_1\nonumber\\
&-b_1(1+\lambda^{-1}_{S,2}d^{(k)}_1-(\lambda^{(k)}_{S,1})^{-1}d^{(k)}_1)^{-2}\nonumber\\
&+c(\lambda^{(k)}_{X,1})^2(\lambda^{(k)}_{S,1})^{-2}=0,\label{eq:kkt1}\\
&a_2-b_2+c(k-1)\lambda^2_{X,2}\lambda^{-2}_{S,2}=0,\label{eq:kkt2}\\
&-\frac{1}{2\delta}+b_1+b_2=0,\label{eq:kkt3}\\
&a_1(d^{(k)}_1-\lambda^{(k)}_{S,1})=0,\label{eq:kkt4}\\
&a_2(d^{(k)}_2-\lambda_{S,2})=0,,\label{eq:kkt5}\\
&b_1(\delta-((d^{(k)}_1)^{-1}+\lambda^{-1}_{S,2}-(\lambda^{(k)}_{S,1})^{-1})^{-1})=0,\label{eq:kkt6}\\
&b_2(\delta-d^{(k)}_2)=0,\label{eq:kkt7}\\
&c((\lambda^{(k)}_{X,1})^2(\lambda^{(k)}_{S,1})^{-2}d^{(k)}_1+\lambda^{(k)}_{X,1}-(\lambda^{(k)}_{X,1})^2(\lambda^{(k)}_{S,1})^{-1}\nonumber\\
&+(k-1)(\lambda^2_{X,2}\lambda^{-2}_{S,2}d^{(k)}_2+\lambda_{X,2}-\lambda^2_{X,2}\lambda^{-1}_{S,2})-k d_k)\nonumber\\
&=0.\label{eq:kkt8}
\end{align}
Assume $d_k\in(d^{(k)}_{\min},\gamma_X)$. It can be verified via algebraic manipulations that $\eta(d^{(k)}_1,d^{(k)}_2,\delta)=\overline{r}(d_k)$ for
\begin{align}
&d^{(k)}_1\triangleq((\lambda^{(k)}_{S,1})^{-1}+(\lambda^{(k)}_Q)^{-1})^{-1},\nonumber\\
&d^{(k)}_2\triangleq(\lambda^{-1}_{S,2}+(\lambda^{(k)}_Q)^{-1})^{-1},\nonumber\\
&\delta\triangleq(\lambda^{-1}_{S,2}+(\lambda^{(k)}_Q)^{-1})^{-1},\label{eq:uniquedelta}
\end{align}
where $\lambda^{(k)}_Q$ is given by (\ref{eq:lambdaQ}). We shall identify the condition under which this specific $(d^{(k)}_1,d^{(k)}_2,\delta)$ is a minimizer of ($\mathbf{P}$). Clearly, (\ref{eq:kkt6})--(\ref{eq:kkt8}) are satisfied. Moreover, in view of (\ref{eq:kkt4}), (\ref{eq:kkt5}) as well as the fact that $d^{(k)}_1<\lambda^{(k)}_{S,1}$ and $d^{(k)}_2<\lambda_{S,2}$, we must have
\begin{align*}
a_m=0,\quad m=1,2,
\end{align*}
which, together with (\ref{eq:kkt1})--(\ref{eq:kkt3}), implies
\begin{align*}
&b_1=\frac{d^{(k)}_2-d^{(k)}_1+2k c(\lambda^{(k)}_{X,1})^2(\lambda^{(k)}_{S,1})^{-2}(d^{(k)}_1)^2}{2k (d^{(k)}_2)^2},\\
&b_2=(k-1)c\lambda^2_{X,2}\lambda^{-2}_{S,2},\\
&c=\frac{d^{(k)}_1+(k-1)d^{(k)}_2}{(\lambda^{(k)}_{X,1})^2(\lambda^{(k)}_{S,1})^{-2}(d^{(k)}_1)^2+(k-1)\lambda^2_{X,2}\lambda^{-2}_{S,2}(d^{(k)}_2)^2}\\
&\hspace{0.23in}\times\frac{1}{2k}.
\end{align*}
It is obvious that $b_2$ and $c$ are nonnegative. Therefore, it suffices to have $b_1\geq 0$, which is equivalent to condition (\ref{eq:equivcond1}).
Moreover, under this condition, every minimizer $(d^{(k)}_1,d^{(k)}_2,\delta)$ of ($\mathbf{P}$) must satisfy (\ref{eq:uniquedelta}) due to the fact that $\frac{1}{2}\log\frac{\lambda_{S,2}}{\delta}$ is a strictly convex function of $\delta$ (in other words, (\ref{eq:1case1}),
(\ref{eq:equiv2}), (\ref{eq:equiv3}), (\ref{eq:equiv4}), (\ref{eq:2case1}), (\ref{eq:3case1}), and (\ref{eq:equiv7}) imply that $\delta$ is uniquely determined and is given by (\ref{eq:uniquedelta}) when $r=\overline{r}(d_k)$). Hence, under condition (\ref{eq:equivcond1}), when $r=\overline{r}(d_k)$, we can deduce $d_j\geq d^{(k)}_j(d_k)$ by substituting (\ref{eq:uniquedelta}) into (\ref{eq:invoke1}).

When $\lambda_{S,2}\geq\lambda^{(j)}_{S,1}>0$, we can send $\lambda_W\rightarrow\lambda^{(j)}_{S,1}$  and deduce from (\ref{eq:equiv1}), (\ref{eq:equiv5}), (\ref{eq:equiv6}), (\ref{eq:equiv8}), and (\ref{eq:equiv9}) that
\begin{align}
&\hat{\eta}(d^{(k)}_1,d^{(k)}_2,\delta)\leq r,\label{eq:1case1'}\\
&\delta\leq((d^{(k)}_1)^{-1}+(\lambda^{(j)}_{S,1})^{-1}-(\lambda^{(k)}_{S,1})^{-1})^{-1},\label{eq:2case1'}\\
&\delta\leq((d^{(k)}_2)^{-1}+(\lambda^{(j)}_{S,1})^{-1}-\lambda^{-1}_{S,2})^{-1},\label{eq:3case1'}\\
&\delta\leq d^{(j)}_1,\label{eq:4case1'}\\
&\delta\leq ((d^{(j)}_2)^{-1}+(\lambda^{(j)}_{S,1})^{-1}-\lambda^{-1}_{S,2})^{-1},\label{eq:5case1'}
\end{align}
where
\begin{align*}
&\hat{\eta}(d^{(k)}_1,d^{(k)}_2,\delta)\\
&\triangleq\frac{1}{2k}\log\frac{(\lambda^{(k)}_{S,1})^2}{(\lambda^{(k)}_{S,1}-\lambda^{(j)}_{S,1})d^{(k)}_1+\lambda^{(k)}_{S,1}\lambda^{(j)}_{S,1}}\nonumber\\
&+\frac{k-1}{2k}\log\frac{\lambda^2_{S,2}}{(\lambda_{S,2}-\lambda^{(j)}_{S,1})d^{(k)}_2+\lambda_{S,2}\lambda^{(j)}_{S,1}}+\frac{1}{2}\log\frac{\lambda^{(j)}_{S,1}}{\delta}.
\end{align*}
Furthermore, combining (\ref{eq:equiv10}), (\ref{eq:4case1'}), and (\ref{eq:5case1'}) gives
\begin{align}
d_j&\geq\frac{1}{j}((\lambda^{(j)}_{X,1})^{2}(\lambda^{(j)}_{S,1})^{-2}\delta+\lambda^{(j)}_{X,1}-(\lambda^{(j)}_{X,1})^2(\lambda^{(j)}_{S,1})^{-1})\nonumber\\
&\quad+\frac{j-1}{j}(\lambda^2_{X,2}\lambda^{-2}_{S,2}(\delta^{-1}+\lambda^{-1}_{S,2}-(\lambda^{(j)}_{S,1})^{-1})^{-1}\nonumber\\
&\qquad+\lambda_{X,2}-\lambda^2_{X,2}\lambda^{-1}_{S,2}).\label{eq:invoke2}
\end{align}
Now consider the following convex optimization problem:
\begin{align*}
\min\limits_{d^{(k)}_1,d^{(k)}_2,\delta}\hat{\eta}(d^{(k)}_1,d^{(k)}_2,\delta)\qquad\qquad\mbox{(} \hat{\mathbf{P}} \mbox{)}
\end{align*}
subject to (\ref{eq:equiv2}), (\ref{eq:equiv3}), (\ref{eq:equiv4}), (\ref{eq:2case1'}), (\ref{eq:3case1'}), and (\ref{eq:equiv7}).
According to the Karush-Kuhn-Tucker conditions, $(d^{(k)}_{1},d^{(k)}_2,\delta)$ is a minimizer of the convex optimization problem ($\hat{\mathbf{P}}$) if and only if (\ref{eq:equiv2}), (\ref{eq:equiv3}), (\ref{eq:equiv4}), (\ref{eq:2case1'}), (\ref{eq:3case1'}), and (\ref{eq:equiv7}) are satisfied, and there exist nonnegative $\hat{a}_1,\hat{a}_2,\hat{b}_1,\hat{b}_2,\hat{c}$ such that
\begin{align}
&\frac{\lambda^{(j)}_{S,1}-\lambda^{(k)}_{S,1}}{2k ((\lambda^{(k)}_{S,1}-\lambda^{(j)}_{S,1})d^{(k)}_1+\lambda^{(k)}_{S,1}\lambda^{(j)}_{S,1})}+\hat{a}_1\nonumber\\
&-\hat{b}_1(1+(\lambda^{(j)}_{S,1})^{-1}d^{(k)}_1-(\lambda^{(k)}_{S,1})^{-1}d^{(k)}_1)^{-2}\nonumber\\
&+\hat{c}(\lambda^{(k)}_{X,1})^2(\lambda^{(k)}_{S,1})^{-2}=0,\label{eq:kkt1'}\\
&\frac{(k-1)(\lambda^{(j)}_{S,1}-\lambda_{S,2})}{2k((\lambda_{S,2}-\lambda^{(j)}_{S,1})d^{(k)}_2+\lambda_{S,2}\lambda^{(j)}_{S,1})}+\hat{a}_2\nonumber\\
&-\hat{b}_2(1+(\lambda^{(j)}_{S,1})^{-1}d^{(k)}_2-\lambda^{-1}_{S,2}d^{(k)}_2)^{-2}\nonumber\\
&+\hat{c}(k-1)\lambda^2_{X,2}\lambda^{-2}_{S,2}=0,\label{eq:kkt2'}\\
&-\frac{1}{2\delta}+\hat{b}_1+\hat{b}_2=0,\label{eq:kkt3'}\\
&\hat{a}_1(d^{(k)}_1-\lambda^{(k)}_{S,1})=0,\label{eq:kkt4'}\\
&\hat{a}_2(d^{(k)}_2-\lambda_{S,2})=0,,\label{eq:kkt5'}\\
&\hat{b}_1(\delta-((d^{(k)}_1)^{-1}+(\lambda^{(j)}_{S,1})^{-1}-(\lambda^{(k)}_{S,1})^{-1})^{-1})=0,\label{eq:kkt6'}\\
&\hat{b}_2(\delta-((d^{(k)}_2)^{-1}+(\lambda^{(j)}_{S,1})^{-1}-\lambda^{-1}_{S,2})^{-1})=0,\label{eq:kkt7'}\\
&\hat{c}((\lambda^{(k)}_{X,1})^2(\lambda^{(k)}_{S,1})^{-2}d^{(k)}_1+\lambda^{(k)}_{X,1}-(\lambda^{(k)}_{X,1})^2(\lambda^{(k)}_{S,1})^{-1}\nonumber\\
&+(k-1)(\lambda^2_{X,2}\lambda^{-2}_{S,2}d^{(k)}_2+\lambda_{X,2}-\lambda^2_{X,2}\lambda^{-1}_{S,2})-k d_k)\nonumber\\
&=0.\label{eq:kkt8'}
\end{align}
Assume $d_k\in(d^{(k)}_{\min},\gamma_X)$. It can be verified via algebraic manipulations that $\hat{\eta}(d^{(k)}_1,d^{(k)}_2,\delta)=\overline{r}(d_k)$ for
\begin{align}
&d^{(k)}_1\triangleq((\lambda^{(k)}_{S,1})^{-1}+(\lambda^{(k)}_Q)^{-1})^{-1},\nonumber\\
&d^{(k)}_2\triangleq(\lambda^{-1}_{S,2}+(\lambda^{(k)}_Q)^{-1})^{-1},\nonumber\\
&\delta\triangleq((\lambda^{(j)}_{S,1})^{-1}+(\lambda^{(k)}_Q)^{-1})^{-1},\label{eq:uniquedelta'}
\end{align}
where $\lambda^{(k)}_Q$ is given by (\ref{eq:lambdaQ}). We shall identify the conditions under which this specific $(d^{(k)}_1,d^{(k)}_2,\delta)$ is a minimizer of ($\hat{\mathbf{P}}$).  Clearly, (\ref{eq:kkt6'})--(\ref{eq:kkt8'}) are satisfied. Moreover, in view of (\ref{eq:kkt4'}), (\ref{eq:kkt5'}) as well as the fact that $d^{(k)}_1<\lambda^{(k)}_{S,1}$ and $d^{(k)}_2<\lambda_{S,2}$, we must have
\begin{align*}
\hat{a}_m=0,\quad m=1,2,
\end{align*}
which, together with (\ref{eq:kkt1'})--(\ref{eq:kkt3'}), implies
\begin{align*}
&\hat{b}_1=\frac{\delta-d^{(k)}_1+2k\hat{c}(\lambda^{(k)}_{X,1})^2(\lambda^{(k)}_{S,1})^{-2}(d^{(k)}_1)^2}{2k \delta^2},\\
&\hat{b}_2=\frac{(k-1)(\delta-d^{(k)}_2)+2k(k-1)\hat{c}\lambda^2_{X,2}\lambda^{-2}_{S,2}(d^{(k)}_2)^2}{2k\delta^2},\\
&\hat{c}=\frac{d^{(k)}_1+(k-1)d^{(k)}_2}{(\lambda^{(k)}_{X,1})^2(\lambda^{(k)}_{S,1})^{-2}(d^{(k)}_1)^2+(k-1)\lambda^2_{X,2}\lambda^{-2}_{S,2}(d^{(k)}_2)^2}\\
&\hspace{0.23in}\times\frac{1}{2k}.
\end{align*}
It is obvious that $\hat{c}$ is nonnegative. Therefore, it suffices to have $\hat{b}_1\geq 0$ and $\hat{b}_2\geq 0$, which are equivalent to conditions (\ref{eq:equivcond3}) and (\ref{eq:equivcond4}), respectively (note that, when $j=k$, condition (\ref{eq:equivcond3}) is redundant and condition (\ref{eq:equivcond4}) is simplified to condition (\ref{eq:equivcond2})).
Moreover, under these conditions, every minimizer $(d^{(k)}_1,d^{(k)}_2,\delta)$ of ($\hat{\mathbf{P}}$) must satisfy (\ref{eq:uniquedelta'}) due to the fact that $\frac{1}{2}\log\frac{\lambda^{(j)}_{S,1}}{\delta}$ is a strictly convex function of $\delta$ (in other words, (\ref{eq:1case1'}),
(\ref{eq:equiv2}), (\ref{eq:equiv3}), (\ref{eq:equiv4}), (\ref{eq:2case1'}), (\ref{eq:3case1'}), and (\ref{eq:equiv7}) imply that $\delta$ is uniquely determined and is given by (\ref{eq:uniquedelta'}) when $r=\overline{r}(d_k)$). Hence, under conditions (\ref{eq:equivcond3}) and (\ref{eq:equivcond4}), when $r=\overline{r}(d_k)$, we can deduce $d_j\geq d^{(k)}_j(d_k)$ by substituting (\ref{eq:uniquedelta'}) into (\ref{eq:invoke2}).

For the degenerate case $\lambda^{(j)}_{S,1}>\lambda_{S,2}=0$, we have
\begin{align*}
&\overline{r}^{(k)}(d_k)=\frac{1}{2k}\log\frac{\gamma^2_X}{\gamma_Sd_k-\gamma_X\gamma_Z},\\
&d^{(k)}_j(d_k)=\frac{(j-k)\gamma^2_X\gamma_Z+(k\gamma_S-j\gamma_Z)\gamma_Xd_k}{(j\gamma_S-k\gamma_Z)\gamma_X-(j-k)\gamma_Sd_k}.
\end{align*}
The desired conclusion that $r\geq \overline{r}^{(k)}(d_k)$ and that $d_j\geq d^{(k)}_j(d_k)$ when $r=\overline{r}^{(k)}(d_k)$ follows from the corresponding result for the quadratic Gaussian multiple description problem \cite{WV09,SSC14}. Note that $(k-1)\lambda^2_{X,2}(\lambda^{(k)}_{S,1})\mu^{(k)}(\mu^{(k)}-1)+k(\lambda^{(k)}_{X,1})^2\lambda^2_{S,2}=0$ (consequently, condition (\ref{eq:equivcond1}) is satisfied) when $\lambda^{(j)}_{S,1}>\lambda_{S,2}=0$. Finally, consider the degenerate case  $\lambda_{S,2}>\lambda^{(\ell)}_{S,1}=0$. It can be verified that
\begin{align*}
\overline{r}^{(\ell)}(d_{\ell})=\frac{\ell-1}{2\ell}\log\frac{(\ell-1)\lambda^2_{X,2}}{\ell\lambda_{S,2}d_{\ell}-(\ell-1)\lambda_{X,2}\lambda_{Z,2}},
\end{align*}
which coincides with the rate-distortion function (normalized by $\ell$) of the corresponding centralized remote source coding problem. Therefore, we must have $r\geq\overline{r}^{(\ell)}(d_{\ell})$.
Also, note that $(\lambda^{(\ell)}_{X,1})^2\lambda^2_{S,2}\nu^{(\ell)}(\nu^{(\ell)}-1)+\ell\lambda^2_{X,2}(\lambda^{(\ell)}_{S,1})^2=0$ (consequently, condition (\ref{eq:equivcond2}) is satisfied for $k=\ell$) when $\lambda_{S,2}>\lambda^{(\ell)}_{S,1}=0$. This completes the proof of Theorem \ref{thm:theorem2}.




\section{Conclusion}\label{sec:conclusion}

We have studied the problem of robust distributed compression of correlated Gaussian sources in a symmetric setting and obtained a characterization of certain extremal points of the rate-distortion region. It is expected that one can make further progress by integrating our techniques with those developed for the quadratic Gaussian multiple description problem.

\appendices
\section{Calculation of $d^{(j)}_{\min}$}\label{app:dmin}

Assuming $\Gamma^{(j)}_S\succ 0$ (i.e., $\lambda^{(j)}_{S,1}>0$ and $\lambda_{S,2}>0$), we have
\begin{align*}
&\sum_{i=1}^j\mathbb{E}[(X_i-\mathbb{E}[X_i|S_1,\cdots,S_j])^2]]\\
&=\mathrm{tr}(\Gamma^{(j)}_X-\Gamma^{(j)}_X(\Gamma^{(j)}_S)^{-1}\Gamma^{(j)}_X)\\
&=\mathrm{tr}(\Lambda^{(j)}_X-\Lambda^{(j)}_X(\Lambda^{(j)}_S)^{-1}\Lambda^{(j)}_X)\\
&=\frac{\lambda^{(j)}_{X,1}\lambda^{(j)}_{Z,1}}{\lambda^{(j)}_{S,1}}+(j-1)\frac{\lambda_{X,2}\lambda_{Z,2}}{\lambda_{S,2}},
\end{align*}
from which the desired result follows immediately. The degenerate case $\lambda^{(j)}_{S,1}=0$ or $\lambda_{S,2}=0$ can be handled by performing the above analysis in a suitable subspace.
\section{Proof of (\ref{eq:sigma}) and (\ref{eq:delta1})}\label{app:constraints}

For $t=1,\cdots,n$,
\begin{align*}
(G_{i,\mathcal{A}}(t))^T_{i\in\mathcal{A}}&\triangleq (U_i(t))^T_{i\in\mathcal{A}}-\mathbb{E}[(U_i(t))^T_{i\in\mathcal{A}}|(S_{i}(t))^T_{i\in\mathcal{A}}]\\
&=(U_i(t))^T_{i\in\mathcal{A}}-\Gamma^{(k)}_U(\Gamma^{(k)}_S)^{-1}(S_i(t))^T_{i\in\mathcal{A}},
\end{align*}
which is an $k$-dimensional zero-mean Gaussian random vector with covariance $\Gamma^{(k)}_U-\Gamma^{(k)}_U(\Gamma^{(k)}_S)^{-1}\Gamma^{(k)}_U$ and is independent of $(S^n_{i})^T_{i\in\mathcal{A}}$. As a consequence,
\begin{align*}
&(\hat{U}_{i,\mathcal{A}}(t))^T_{i\in\mathcal{A}}=\Gamma^{(k)}_U(\Gamma^{(k)}_S)^{-1}(\hat{S}_{i,\mathcal{A}}(t))^T_{i\in\mathcal{A}},\\
&\hspace{2in} t=1,\cdots,n.
\end{align*}
Now it can be readily verified that
\begin{align*}
\Sigma_{\mathcal{A}}(t)&=\Gamma^{(k)}_U(\Gamma^{(k)}_S)^{-1}D_{\mathcal{A}}(t)(\Gamma^{(k)}_S)^{-1}\Gamma^{(k)}_U\\
&\quad+\mathbb{E}[(G_{i,\mathcal{A}}(t))^T_{i\in\mathcal{A}}(G_{i,\mathcal{A}}(t))_{i\in\mathcal{A}}]\\
&=\Gamma^{(k)}_U(\Gamma^{(k)}_S)^{-1}D_{\mathcal{A}}(t)(\Gamma^{(k)}_S)^{-1}\Gamma^{(k)}_U+\Gamma^{(k)}_U\\
&\quad-\Gamma^{(k)}_U(\Gamma^{(k)}_S)^{-1}\Gamma^{(k)}_U,\quad t=1,\cdots,n,
\end{align*}
from which (\ref{eq:sigma}) follows immediately.

For $t=1,\cdots,n$, we have
\begin{align}
\Delta_{\mathcal{A}}(t)&\preceq\mathbb{E}[(S_i(t)-\tilde{S}'_{i,\mathcal{A}}(t))^T_{i\in\mathcal{A}}(S_i(t)-\tilde{S}'_{i,\mathcal{A}}(t))_{i\in\mathcal{A}}]\nonumber\\
&=((D_{\mathcal{A}}(t))^{-1}+(\Lambda^{(k)}_W)^{-1}-(\Gamma^{(k)}_S)^{-1})^{-1},\label{eq:linearMMSE}
\end{align}
where $(\tilde{S}'_{i,\mathcal{A}}(t))^T_{i\in\mathcal{A}}$ denotes the linear MMSE estimator of $(S_i(t))^T_{i\in\mathcal{A}}$ based on $(\hat{S}_{i,\mathcal{A}}(t))^T_{i\in\mathcal{A}}$ and $(U_i(t))^T_{i\in\mathcal{A}}$. Since $(A^{-1}+B^{-1})^{-1}$ is matrix concave in $A$ for $A\succ 0$ and $B\succ 0$, it follows that
\begin{align}
&\frac{1}{n}\sum\limits_{t=1}^n((D_{\mathcal{A}}(t))^{-1}+(\Lambda^{(k)}_W)^{-1}-(\Gamma^{(k)}_S)^{-1})^{-1}\nonumber\\
&\preceq(D^{-1}_{\mathcal{A}}+(\Lambda^{(k)}_W)^{-1}-(\Gamma^{(k)}_S)^{-1})^{-1}.\label{eq:matrixconcave}
\end{align}
Combing (\ref{eq:linearMMSE}) and (\ref{eq:matrixconcave}) proves (\ref{eq:delta1}).



\ifCLASSOPTIONcaptionsoff
  \newpage
\fi


\begin{thebibliography}{1}

\bibitem{SW73}
D. Slepian and J. K. Wolf, ``Noiseless coding of correlated information sources," {\em IEEE  Trans.  Inf. Theory}, vol. IT-19, no. 4, pp. 471–-480, Jul. 1973.



\bibitem{Oohama97}
Y. Oohama,  ``Gaussian  multiterminal  source  coding," {\em IEEE  Trans.  Inf. Theory}, vol. 43, no. 6, pp. 1912--1923,  Nov. 1997.

\bibitem{Oohama98}
Y. Oohama, ``The rate-distortion function for the quadratic Gaussian CEO problem," {\em IEEE  Trans.  Inf. Theory}, vol. 44, no. 3, pp. 1057–-1070, May 1998.


\bibitem{PTR04}
V. Prabhakaran, D. Tse, and K. Ramchandran, ``Rate region of the quadratic Gaussian CEO problem," in {\em Proc.  IEEE  Int.  Symp.  Inform.
Theory (ISIT)}, Chicago,  IL, USA, Jun./Jul. 2004, p. 117.

\bibitem{CZBW04}
J.  Chen,  X.  Zhang,  T.  Berger,  and  S.  B.  Wicker,  ``An  upper  bound  on the  sum-rate  distortion  function  and  its  corresponding  rate  allocation
schemes  for  the CEO problem," {\em IEEE J.  Sel.  Areas Commun.},  vol. 22, no. 6, pp. 977--987, Aug. 2004.


\bibitem{Oohama05}
Y. Oohama, ``Rate-distortion theory for Gaussian multiterminal source coding systems with several side informations at the decoder," {\em IEEE
Trans. Inf. Theory}, vol. 51, no. 7, pp. 2577–2593, Jul. 2005.

\bibitem{CB08}
J. Chen and T. Berger, ``Successive Wyner-Ziv coding scheme and its application to the quadratic Gaussian
  CEO problem," {\em IEEE  Trans.  Inf.  Theory}, vol. 54, no. 4, pp. 1586--1603, Apr. 2008.

\bibitem{WTV08}
A. B. Wagner,  S. Tavildar, and P. Viswanath,  ``Rate  region  of  the quadratic  Gaussian  two-encoder  source-coding  problem," {\em IEEE  Trans. Inf. Theory}, vol. 54, no. 5, pp. 1938--1961,  May 2008.

\bibitem{TVW10}
S.  Tavildar,  P.  Viswanath,  and  A.  B.  Wagner,  ``The  Gaussian  many-help-one  distributed  source  coding  problem," {\em IEEE Trans.  Inf.  Theory},
vol. 56, no. 1, pp. 564–-581,  Jan. 2010.




\bibitem{WCW10}
J. Wang, J. Chen, and X. Wu, ``On  the  sum  rate  of  Gaussian  multiterminal source coding: New proofs and results," {\em IEEE Trans. Inf. Theory},
vol. 56, no. 8, pp. 3946--3960, Aug. 2010.

\bibitem{YX12}
Y. Yang and Z. Xiong, ``On the generalized Gaussian CEO problem," {\em IEEE Trans. Inf. Theory}, vol. 58, no. 6, pp. 3350--3372, Jun. 2012.

\bibitem{YZX13}
Y. Yang, Y. Zhang, and Z. Xiong, ``A new sufficient condition  for sum-rate tightness in quadratic Gaussian multiterminal source coding," {\em IEEE Trans. Inf. Theory}, vol. 59, no. 1, pp. 408--423,  Jan. 2013.


\bibitem{WC13}
J. Wang and  J. Chen,  ``Vector  Gaussian  two-terminal  source  coding," {\em IEEE Trans. Inf. Theory}, vol. 59, no. 6, pp. 3693--3708,  Jun. 2013.



\bibitem{WC14}
J. Wang and  J. Chen,  ``Vector  Gaussian  multiterminal  source  coding," {\em IEEE Trans. Inf. Theory}, vol. 60, no. 9, pp. 5533--5552,  Sep. 2014.

\bibitem{Oohama14}
Y. Oohama, ``Indirect and direct Gaussian distributed source coding problems," {\em IEEE Trans. Inf. Theory}, vol. 60, no. 12, pp. 7506--7539, Dec. 2014.


\bibitem{CEK17}
J. Chen, F. Etezadi, and A. Khisti, ``Generalized Gaussian multiterminal source coding and probabilistic graphical models," in
{\em Proc.  IEEE  Int.  Symp.  Inform.
Theory (ISIT)},  Aachen, Germany, Jun. 25 - 30, 2017,  pp. 719--723.


\bibitem{CXCWW17}
J. Chen, L. Xie, Y. Chang, J. Wang, and Y. Wang, ``Generalized Gaussian multiterminal source coding: The symmetric case," arXiv:1710.04750.







\bibitem{Ozarow80}
L. Ozarow,  ``On  a  source-coding  problem  with  two  channels  and  three
receivers," {\em Bell Syst. Tech. J.}, vol. 59, no. 10, pp. 1909--1921, Dec. 1980.

\bibitem{EGC82}
A.  A.  El  Gamal  and  T.  M.  Cover,  ``Achievable   rates  for  multiple descriptions," {\em IEEE  Trans.  Inf.  Theory},  vol.  28,  no.  6,  pp.  851--857, Nov. 1982.


\bibitem{VKG03}
R.  Venkataramani,  G.  Kramer,  and  V.  K.  Goyal,  ``Multiple  description coding  with  many  channels," {\em IEEE  Trans.  Inf.  Theory},  vol.  49,  no.  9, pp. 2106--2114,  Sep. 2003.

\bibitem{PPR04}
S.  S.  Pradhan,  R.  Puri,  and  K.  Ramchandran,  ``$n$-channel  symmetric multiple   descriptions---Part   I: $(n,k)$ source-channel   erasure   codes," {\em IEEE Trans. Inf. Theory}, vol. 50, no. 1, pp. 47--61, Jan. 2004.

\bibitem{PPR05}
R.  Puri,  S.  S.  Pradhan,  and  K.  Ramchandran,  ``$n$-channel  symmetric multiple  descriptions---Part  II:  An  achievable  rate-distortion  region," {\em IEEE Trans. Inf. Theory}, vol. 51, no. 4, pp. 1377--1392,  Apr. 2005.

\bibitem{CTBH06}
J. Chen, C. Tian, T. Berger, and S. S. Hemami, ``Multiple description quantization via Gram-Schmidt orthogonalization,"
    {\em IEEE  Trans.  Inf.  Theory}, vol. 52, no. 12, pp. 5197--5217, Dec. 2006.



\bibitem{WV07}
H. Wang and P. Viswanath,  ``Vector Gaussian multiple description  with individual and central receivers," {\em IEEE Trans. Inf. Theory}, vol. 53, no. 6, pp. 2133--2153,  Jun. 2007.

\bibitem{TCD08}
C. Tian, J. Chen, and S. Diggavi, ``Multiuser successive refinement and multiple description coding," {\em IEEE  Trans.  Inf.  Theory}, vol. 54, no. 2, pp. 921--931, Feb. 2008.

\bibitem{WV09}
H. Wang and P. Viswanath,  ``Vector Gaussian multiple description  with two  levels   of  receivers," {\em IEEE  Trans.  Inf.  Theory},   vol.  55,  no.  1, pp. 401--410,  Jan. 2009.

\bibitem{CTD09}
J. Chen, C. Tian, and S. Diggavi, ``Multiple description coding for stationary Gaussian sources," {\em IEEE  Trans.  Inf.  Theory}, vol. 55, no. 6, pp. 2868--2881, Jun. 2009.

\bibitem{Chen09}
J.  Chen,  ``Rate  region  of  Gaussian  multiple  description  coding  with individual  and  central  distortion  constraints," {\em IEEE  Trans.  Inf.  Theory}, vol. 55, no. 9, pp. 3991--4005,  Sep. 2009.



\bibitem{TC10}
C.   Tian   and   J.  Chen,   ``New   coding   schemes   for   the   symmetric $K$-description   problem," {\em IEEE  Trans.  Inf.  Theory},  vol.  56,  no.  10, pp. 5344--5365,  Oct. 2010.

\bibitem{CDZW10}
J. Chen, S. Dumitrescu, Y. Zhang, and J. Wang, ``Robust multiresolution coding," {\em IEEE Trans. Commun.},  vol. 58, no. 11, pp. 3186--3195, Nov. 2010.

\bibitem{WCZCP11}
J. Wang, J. Chen, L. Zhao, P. Cuff, and H. Permuter, ``On the role of the refinement layer in multiple description coding and scalable coding," {\em IEEE  Trans.  Inf.  Theory}, vol. 57, no. 3, pp. 1443--1456, Mar. 2011.

\bibitem{CZD12}
J. Chen, Y. Zhang, and S. Dumitrescu, ``Gaussian multiple description coding with low-density generator matrix codes," {\em IEEE Trans. Commun.}, vol. 60, no. 3, pp. 676--687, Mar. 2012.

\bibitem{ZDCS12}
Y. Zhang, S. Dumitrescu, J. Chen, and Z. Sun, ``LDGM-based multiple description coding for finite alphabet sources,"  {\em IEEE Trans. Commun.}, vol. 60, no. 12, pp. 3671--3682, Dec. 2012.

\bibitem{FWSC13}
Y. Fan, J. Wang, J. Sun, and J. Chen, ``On the generalization of natural type
selection to multiple description coding," {\em IEEE Trans. Commun.}, vol. 61, pp. 1361--1373, Apr. 2013.



\bibitem{SSC14}
L. Song, S. Shao, and J. Chen, ``A lower bound on the sum rate of multiple description coding with symmetric distortion constraints," {\em IEEE  Trans.  Inf.  Theory}, vol. 60, no. 12, pp. 7547--7567,  Dec. 2014.

\bibitem{XCW17}
Y. Xu, J. Chen, and Q. Wang, ``The sum rate of vector Gaussian multiple description coding with tree-structured covariance distortion constraints," {\em IEEE  Trans.  Inf.  Theory},  vol. 63, no. 10, pp. 6547--6560, Oct. 2017.



\bibitem{IPRP05}
P.  Ishwar,  R. Puri,  K. Ramchandran,  and  S.  S.  Pradhan,  ``On rate-constrained distributed estimation in
unreliable sensor networks," {\em IEEE J.  Sel.  Areas Commun.},  vol. 23, no. 4, pp. 765--775, Apr. 2005.




\bibitem{CB082}
J. Chen and T. Berger, ``Robust distributed source coding,"  {\em IEEE
Trans. Inf. Theory}, vol. 54, no. 8, pp. 3385--3398,  Aug. 2008.


\bibitem{CW08}
J. Chen and A. B. Wagner, ``A semicontinuity theorem and its application to network source coding,"   in
{\em Proc.  IEEE  Int.  Symp.  Inform. Theory (ISIT)},   Toronto, Canada, Jul. 6 - 11, 2008, pp. 429--433.












\bibitem{Berger78}
T.  Berger,  ``Multiterminal   source  coding,"  in {\em The  Information  Theory   Approach   to   Communications}
(CISM   International   Centre   for
Mechanical  Sciences),  vol. 229,  G.  Longo,  Ed.  New  York,  NY,  USA:
Springer-Verlag,  1978, pp. 171--231.

\bibitem{Tung78}
S.-Y.  Tung,  ``Multiterminal  source  coding,"  Ph.D.  dissertation,  School
Electr.  Eng., Cornell  Univ., Ithaca,  NY, USA, 1978.





\bibitem{XCWB07}
X. Zhang, J. Chen, S. B. Wicker, and T. Berger, ``Successive coding in multiuser information theory," {\em IEEE Trans. Inf. Theory}, vol. 53, no. 6, pp. 2246--2254, Jun. 2007.

\bibitem{WA08}
A. B. Wagner and V. Anantharam,  ``An  improved  outer  bound  for
multiterminal  source  coding," {\em IEEE  Trans. Inf. Theory}, vol. 54, no. 5, pp. 1919--1937,  May 2008.


\bibitem{EGK11}
A. El Gamal and Y.-H. Kim, {\em Network Information Theory}. Cambridge, U.K.: Cambridge University Press, 2011.


\end{thebibliography}
\end{document}